\documentclass[fleqn,usenatbib]{mnras}

\usepackage{newtxtext,newtxmath}

\usepackage[T1]{fontenc}

\DeclareRobustCommand{\VAN}[3]{#2}
\let\VANthebibliography\thebibliography
\def\thebibliography{\DeclareRobustCommand{\VAN}[3]{##3}\VANthebibliography}

\usepackage{graphicx}	
\usepackage{amsmath}	

\title[Probing 3D Magnetic Field with Dust Polarization]{Probing Three-Dimensional Magnetic Fields: I - Polarized Dust Emission}

\author[Hu \& Lazarian]{
Yue Hu$^{1,2}$\thanks{E-mail: yue.hu@wisc.edu},
A. Lazarian$^{2,3}$\thanks{E-mail: alazarian@facstaff.wisc.edu}
\\
$^{1}$Department of Physics, University of Wisconsin-Madison, Madison, WI, 53706, USA\\
$^{2}$Department of Astronomy, University of Wisconsin-Madison, Madison, WI, 53706, USA\\
$^{3}$Centro de Investigación en Astronomía, Universidad Bernardo O’Higgins, Santiago, General Gana 1760, 8370993,
Chile\\
}

\date{Accepted XXX. Received YYY; in original form ZZZ}

\pubyear{2022}

\begin{document}
\label{firstpage}
\pagerange{\pageref{firstpage}--\pageref{lastpage}}
\maketitle

\begin{abstract}
Polarized dust emission is widely used to trace the plane-of-the-sky (POS) component of interstellar magnetic fields in two dimensions. Its potential to access three-dimensional magnetic fields, including the inclination angle of the magnetic fields relative to the line-of-sight (LOS), is crucial for a variety of astrophysical problems. Based on the statistical features of observed polarization fraction and POS Alfv\'en Mach number $\overline{M_{\rm A}}_{,\bot}$ distribution, we present a new method for estimating the inclination angle. The magnetic field fluctuations raised by anisotropic magnetohydrodynamic (MHD) turbulence are taken into account in our method. By using synthetic dust emission generated from 3D compressible MHD turbulence simulations, we show that the fluctuations are preferentially perpendicular to the mean magnetic field. We find the inclination angle is the major agent for depolarization, while fluctuations of magnetic field strength and density have an insignificant contribution. We propose and demonstrate that the mean inclination angle over a region of interest can be calculated from the polarization fraction in a strongly magnetized reference position, where $\overline{M_{\rm A}}_{,\bot}^2\ll1$. We test and show that the new method can trace the 3D magnetic fields in sub-Alfv\'enic, trans-Alfv\'enic, and moderately super-Alfv\'enic conditions ($0.4\lesssim M_{\rm A}\lesssim1.2$). We numerically quantify that the difference of the estimated inclination angle and actual inclination angle ranges from 0 to $20^\circ$ with a median value of $\le10^\circ$.

\end{abstract}




\section{Introduction}
In interstellar medium (ISM), magnetic field is one of the most important components \citep{1988ASSL..133.....R,2016A&A...586A.136P,2016A&A...586A.141P,2017ARA&A..55..111H,2019ApJ...887..136C,HYL20,2022MNRAS.511..829H}. It is crucial in balancing the ISM with gravity \citep{1988ApJ...326L..27M,2003ApJ...599..363A,2018FrASS...5...39W,2020NatAs...4..704A}, regulating turbulent gas flows \citep{1985PASJ...37..515U,2018MNRAS.476..235R,2020NatAs...4.1126B}, and constraining cosmic ray's transport \citep{1966ApJ...146..480J,2011ApJ...741...16G,2013ApJ...779..140X,2021arXiv211115066H,2022arXiv220313952B}. In particular, magnetic field is a key factor influencing the dynamics of the star-forming process in molecular clouds \citep{MK04,Crutcher04,MO07,2012ApJ...757..154L,Crutcher12,2012ApJ...761..156F,HLS21}. In view of its importance, a number of ways to access the magnetic field have been proposed. For instance, polarized dust emission \citep{Lazarian07,BG15,2015A&A...576A.104P,2020A&A...641A..11P,2016ApJ...824..134F,2021MNRAS.tmp.3119L} and synchrotron emission \citep{2008A&A...482..783X,2016A&A...594A..25P,2021ApJ...920....6G} can trace the POS magnetic field, while Zeenman splitting \citep{Crutcher04,Crutcher12} and Faraday rotation \citep{2007ASPC..365..242H,2009ApJ...702.1230T,2012A&A...542A..93O,2016ApJ...824..113X} reveal the LOS magnetic field strength. However, because they probe different regions of the multi-phase ISM, the measurements cannot easily be combined to yield full 3D magnetic field vectors. Probing a three-dimensional magnetic field that includes both the POS and the LOS components simultaneously remains a challenge.

Intense attempts have been undertaken to get the 3D magnetic field at cloud scales. For instance, \citet{2018ApJ...865...59L} proposed a solution using the wavelength derivative of synchrotron polarization. 
\cite{2019A&A...632A..68T,2022arXiv220104718T} used the changed sign of LOS magnetic fields obtained by \cite{2018A&A...614A.100T} to infer bow-shaped magnetic field morphologies across the Orion-A and Perseus molecular clouds. \citet{2020ApJ...902L...7Z} achieved a three-dimensional magnetic field via the fraction and direction of atomic gas's polarization. After that, \citet{HXL21} suggested using MHD turbulence's anisotropic property inherited by young stellar objects to obtain a three-dimensional view of the magnetic field. Similarly, based on anisotropic MHD turbulence, \citet{HLX21a} further extend the method to be applicable for Doppler-shifted emission lines in three-dimension. The LOS and POS components of the magnetic field's orientation and strength can be calculated simultaneously for the latter two methods. 

In addition to the approaches mentioned above, an important step of probing the 3D magnetic field via polarized dust emission was initiated by \citet{2019MNRAS.485.3499C}. The POS magnetic field can be easily inferred from polarization direction based on the fact that dust grains preferentially align with their ambient magnetic fields \citep{Lazarian07,BG15}. To achieve a three-dimensional picture, the inclination angle of the magnetic field relative to the LOS is crucial. As the inclination angle is one of the major agents of depolarizing thermal emission from dust, the polarization fraction intrinsically inherits the angle's information. Therefore, \citet{2019MNRAS.485.3499C} and \cite{2021MNRAS.503.5006S} estimated the inclination angle based on the statistical properties of the observed polarization fraction. Their method assumes an ideal scenario that there are no fluctuations in neither magnetic field's POS nor LOS components. This assumption could be valid for strongly magnetized mediums. However, molecular clouds are typically trans-Alfv\'enic or even super-Alfv\'enic \citep{2016ApJ...832..143F,Hu19a,2021ApJ...913...85H,2021MNRAS.tmp.3119L}, in which the fluctuations are not negligible.

To accommodate the magnetic field fluctuations, here we consider a scenario that the fluctuations arise from anisotropic magnetohydrodynamic (MHD) turbulence based on the fact that molecular cloud is highly turbulent \citep{1981MNRAS.194..809L,1983ApJ...270..105M,1999ARA&A..37..311E,2012A&ARv..20...55H} and is dominated by slow and fast components of MHD turbulence that follow Kolmogorov scaling \citep{2022arXiv220413760Y}. This consideration advantageously simplifies the problem because the most significant fluctuations preferentially appear in the direction perpendicular to the mean magnetic field \citep{GS95,LV99,2003MNRAS.345..325C}. Therefore, we propose a simple model in this work that the local magnetic field along the LOS is built up by a global mean magnetic field and perpendicular fluctuations. This assumption is typically valid for cloud-scale and clump-scale objects in which their magnetic fields' variation along the LOS is insignificant. 

By incorporating the magnetic field fluctuations, this work aims at developing a method to probe the 3D magnetic field in sub-, trans- and super-Alfv\'enic clouds. This method requires the knowledge of the polarization fraction and the POS Alfv\'en Mach number's distributions. The latter can be obtained by a number of approaches. To test the proposed method, we use 3D MHD turbulence simulations to generate synthetic dust emissions. We will show that the assumption of perpendicular fluctuations is also valid in the presence of compressible turbulence.

This paper is organized as follows. We briefly review the basic concepts of MHD turbulence and show the derivation of how to estimate the magnetic field's inclination angle from polarized dust emission. In \S~\ref{sec:data}, we give the details of the simulation's setup and numerical method. We applied our method to the simulations in \S~\ref{sec:result} and made a comparison with the method proposed in \citet{2019MNRAS.485.3499C}. In \S~\ref{sec:dis}, we discuss the systematic uncertainties raised by our assumptions and list several approaches to getting the POS Alfv\'en Mach number's distribution. We summarize our results in \S~\ref{sec:con}.

\section{Theoretical consideration}
\label{sec:theory}
\subsection{Essential elements of MHD turbulence}
Our understanding of MHD turbulence has been significantly changed in the past decades. MHD turbulence was initially considered to be isotropic despite the existence of magnetic fields \citep{1963AZh....40..742I,1965PhFl....8.1385K}.
However, a number of numerical studies \citep{1981PhFl...24..825M,1983JPlPh..29..525S,1984ApJ...285..109H,1965PhFl....8.1385K,1995ApJ...447..706M,2001ApJ...554.1175M,2010ApJ...720..742K,HLX21a} and in situ measurements of solar wind \citep{2016ApJ...816...15W} revealed that the turbulence is anisotropic rather than isotropic when magnetic field's role is not negligible.  

A fundamental work on the anisotropic incompressible MHD turbulence theory was done by \citet{GS95} for the trans-Alfv\'enic regime, i.e. for injection velocity equal to the Alfv\'en velocity and was extended to sub-Alfv\'enic turbulence, i.e. for injection velocity less than the Alfven velocity \citep{LV99}. 

The modern picture of MHD turbulence cascade states that the Alfv\'enic mode cascade is channeled to the field's perpendicular direction. This is achievable because turbulent reconnection, as an intrinsic part of the MHD turbulent cascade, happens over one eddy turnover time and enables the mixing of magnetic field lines perpendicular to the magnetic field direction \citep{LV99}. Thus mixing presents the path of minimal resistance for turbulent motions and the turbulence is channeled along this path. More detail is available in \citet{2019tuma.book.....B}, where the properties of compressible MHD turbulence are described in detail.  

The fluctuations of turbulent velocity therefore is preferentially along the perpendicular direction. From the scaling of MHD turbulence in \citep{LV99} it follows that the ratio of squared velocity fluctuations at scale $l$ along the perpendicular (i.e., $v_{l,\bot}^2$) and the parallel directions (i.e., $v_{l,\parallel}^2$) with respect to the local magnetic field is \citep{HXL21}:

\begin{equation}
\label{eq.vv}
  v_{l,\bot}^2/v_{l,\parallel}^2=(l_\parallel/L_{\rm inj})^{-1/3} M_{\rm A}^{-4/3}, 
\end{equation}
here $M_{\rm A}$ is the Alfv\'en Mach number and $L_{\rm inj}$ is the injection scale of turbulence. $l_{\parallel}$ denotes the scale parallel to the local magnetic field, i.e., the magnetic field passing through the turbulent eddy.\footnote{The notion of local system of reference is fundamental for MHD turbulence scaling. This notion missed in the original study \citep{GS95}, but it naturally follows when turbulent reconnection is considered \citep{LV99}. Numerically, the necessity of using the local system of reference was demonstrated in \cite{2000ApJ...539..273C}.}   The injection scale $L_{\rm inj}$ is approximately 100 pc in our galaxy \citep{1995ApJ...443..209A,2010ApJ...710..853C,2022arXiv220413760Y} and is much greater than the scale $\ll 0.1$~pc that can be resolved in observation \citep{2019ApJ...872..187C,2022A&A...659A..22Z,2022MNRAS.512.1985F}. The velocity fluctuations raised from incompressible MHD turbulence are therefore dominantly along the magnetic field's perpendicular direction for our consideration of molecular clouds and even smaller clumps. 

From the induction equation, one can easily find that the magnetic field fluctuation at scale $l$ is perpendicular to the plane spanned by the global mean magnetic field $\langle\pmb{B}\rangle$ and displacement vector $\pmb{\hat{\xi}}$ of plasma in incompressible turbulence \citep{2003MNRAS.345..325C}:
\begin{equation}
\label{eq.2}
   \delta \pmb{B}_{l,\bot}=\frac{v_{l,\bot}}{v_{\rm A}}({\langle\pmb{B}\rangle}\times\pmb{\hat{\xi}}),
\end{equation}
where $v_{\rm A}$ is the Alfv\'en speed. 
In polarization studies, the statistical description of such fluctuations was provided in \cite{2012ApJ...747....5L}. This agrees well with the numerical studies of compressible MHD turbulence in \cite{HXL21}.  A more detailed discussion of magnetic fluctuations in compressible turbulence is given in \S~\ref{sec:dis}.


\subsection{Estimating inclination angle from dust polarization}
Based on the fact that we in polarization measurements we deal with magnetic field fluctuation  preferentially perpendicular to the mean field, we can investigate the properties of polarized dust emission.

We adopt dust polarization equations from \cite{2015A&A...576A.105P}:
\begin{equation}
\label{eq.dust}
    \begin{aligned}
I(x,y)&=\int n[1-p_0(\sin^2\gamma-2/3)]dz,\\
Q(x,y)&=\int p_0n\frac{B_x^2-B_y^2}{B^2}dz,\\
U(x,y)&=\int p_0n\frac{2B_xB_y}{B^2}dz,\\
\psi(x,y)&=\frac{1}{2}\tan^{-1}(\frac{U}{Q}),
\end{aligned}
\end{equation}
where $n(x,y,z)$ is dust volume density, $\psi$ is polarization angle, and $p_0$ is a polarization fraction parameter related to the intrinsic polarization fraction (assumed to be constant throughout a cloud; \citealt{2019MNRAS.485.3499C}). $B(x,y,z)$ denotes total magnetic field strength, while $B_x(x,y,z)$ and $B_y(x,y,z)$ are its x-axis component and $y$-axis component. $\gamma$ is the magnetic field's inclination angle with respect to the LOS (i.e., the $z$-axis). Accordingly, the polarization fraction is \citep{2000ApJ...544..830F}:
\begin{equation}
\label{eq.4}
\begin{aligned}
        p=\frac{\sqrt{Q^2+U^2}}{I}=p_0\frac{\sqrt{(\int n\frac{B_x^2-B_y^2}{B^2}dz)^2+(\int n\frac{2B_xB_y}{B^2}dz)^2}}{\int ndz-p_0\int n(\sin^2\gamma-2/3)dz}.
\end{aligned}
\end{equation}
\begin{figure}
	\includegraphics[width=1.0\linewidth]{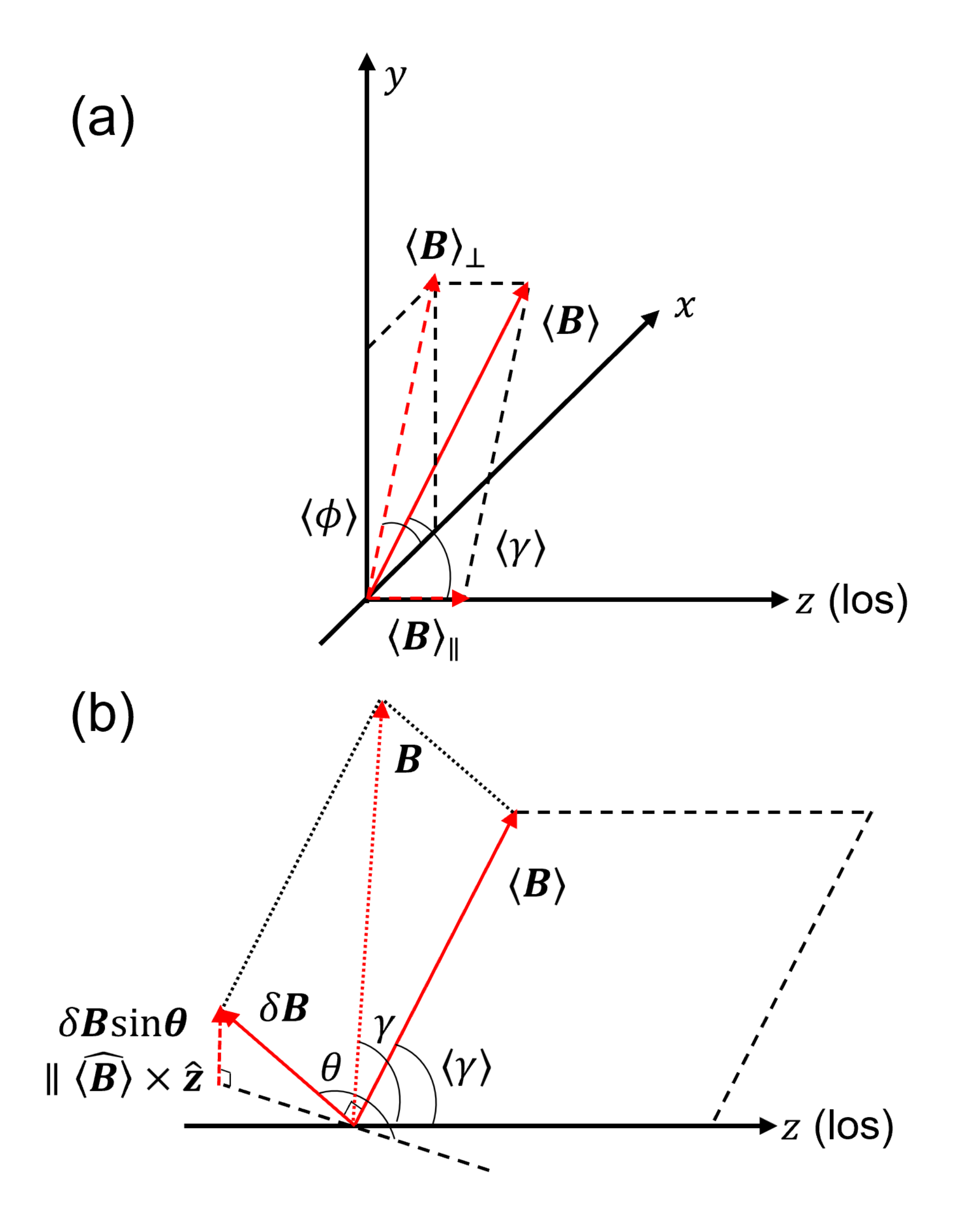}
    \caption{Illustration of the magnetic field configuration. \textbf{Panel a:} configuration of the mean field $\langle\pmb{B}\rangle$. $\langle\pmb{B}\rangle_\bot$ is the magnetic field projected on the POS, i.e. $xy$ plane. $\langle\gamma\rangle$ is the mean inclination angle of the mean magnetic field $\langle\pmb{B}\rangle$ with respect to the LOS. $\langle\phi\rangle=\langle\psi\rangle+\pi/2$ is the magnetic field's angle relative to $x$-axis on the POS. \textbf{Panel b:} configuration of the local total magnetic field $\pmb{B}=\langle\pmb{B}\rangle+\delta\pmb{B}$. The mean field is changed by a perpendicular fluctuation $\delta\pmb{B}$ with an angle $\theta$. Here $\theta$ is the angle between $\delta\pmb{B}$ and the vector (i.e., $\delta\pmb{B}\sin\theta$) that is simultaneously perpendicular to $\langle\pmb{B}\rangle$ and $\hat{\langle\pmb{B}\rangle}\times\hat{\pmb{z}}$. Dashed black lines are within the $\hat{\langle\pmb{B}\rangle}-\hat{\pmb{z}}$ plane, where $\hat{\langle\pmb{B}\rangle}$ and $\hat{z}$ are unit vectors of $\langle\pmb{B}\rangle$ and $\pmb{z}$, respectively.}
    \label{fig:1}
\end{figure}

To describe magnetic field fluctuations, we use a simple configuration of magnetic field (see Fig.~\ref{fig:1}). Assuming the local total magnetic field is built up by a mean magnetic field $\langle \pmb{B}\rangle$ and a fluctuation $\delta \pmb{B}(x,y,z)$:
\begin{equation}
    \pmb{B}(x,y,z)=\langle \pmb{B}\rangle+\delta \pmb{B}(x,y,z),
\end{equation}
The mean-field also has a mean inclination angle $\langle\gamma\rangle$ and POS magnetic field angle $\langle\phi\rangle$. We consider the magnetic field fluctuation $\delta\pmb{B}$ that is preferentially perpendicular to the mean field. However, since $\delta\pmb{B}\propto\hat{\pmb{\xi}}\times\langle \pmb{B}\rangle$, the fluctuation does not necessarily lie on the plane defined by $\langle \pmb{B}\rangle$ and the LOS (i.e., the $z$-axis). Instead, we consider that $\delta\pmb{B}$ has an angle $\theta$ with respect to the $\langle \pmb{B}\rangle-z$ plane. Specifically, $\theta$ is that angle between $\delta\pmb{B}$ and the vector that is simultaneously perpendicular to $\langle\pmb{B}\rangle$ and $\hat{\langle\pmb{B}\rangle}\times\hat{z}$ (see Fig.~\ref{fig:1}). Accordingly, we project the fluctuations and mean field into $x$ and $y$ components:
\begin{equation}
\begin{aligned}
\label{eq:BxBy}
B_x&=\langle B\rangle\sin\langle\gamma\rangle\cos\langle\phi\rangle+(\delta B\cos\theta)\cos\langle\gamma\rangle\cos\langle\phi\rangle-(\delta B\sin\theta)\sin\langle\phi\rangle,\\
B_y&=\langle B\rangle\sin\langle\gamma\rangle\sin\langle\phi\rangle+(\delta B\cos\theta)\cos\langle\gamma\rangle\sin\langle\phi\rangle+(\delta B\sin\theta)\cos\langle\phi\rangle.\\
\end{aligned}
\end{equation}
The first term comes from the mean magnetic field angle $\langle\phi\rangle$ and mean inclination angle $\langle\gamma\rangle$. Their fluctuations $\delta\gamma$ and $\delta\phi$ are introduced by the last two terms involved with $\delta\pmb{B}$.

Note that the direction of $\delta\pmb{B}$ is defined by the displacement vector and the mean field (see Eq.~\ref{eq.2}). As the displacement vector varies in different spatial positions along the LOS, $\theta$ is not a constant. By assuming a uniform distribution of $\theta$ along the LOS, we integrate $\theta$ from 0 to $2\pi$ and take averages:
\begin{equation}
\label{eq:qu}
\begin{aligned}
Q&=\frac{1}{2\pi}\int p_0n\int_0^{2\pi}\frac{B_x^2-B_y^2}{B^2}d\theta dz\\
&=\int p_0n\frac{\cos(2\langle\psi\rangle)[\sin^2\langle\gamma\rangle+ \frac{1}{2}M_{\rm A}^2\cos^2\langle\gamma\rangle-\frac{1}{2} M_{\rm A}^2]}{1+M_{\rm A}^2}dz,\\
U&=\frac{1}{2\pi}\int p_0n\int_0^{2\pi}\frac{2B_xB_y}{B^2}d\theta dz\\
&=\int p_0n\frac{\sin(2\langle\psi\rangle)[\sin^2\langle\gamma\rangle+\frac{1}{2} M_{\rm A}^2\cos^2\langle\gamma\rangle-\frac{1}{2} M_{\rm A}^2]}{1+M_{\rm A}^2}dz,\\ \frac{1}{2\pi}&\int_0^{2\pi}\sin^2\gamma d\theta=\frac{1}{2\pi}\int_0^{2\pi}(1-\cos^2\gamma) d\theta\\
&=1-\frac{M_{\rm A}^2\sin^2\langle\gamma\rangle}{2(M_{\rm A}^2+1)}-\frac{\cos^2\langle\gamma\rangle}{M_{\rm A}^2+1}.
\end{aligned}
\end{equation}

Eq.~\ref{eq:qu} gives the effective values of the three quantities along single LOS. Here $M_{\rm A}=\delta B/\langle B\rangle$ is the  Alfv\'en Mach number.\footnote{For a turbulent volume, the scalar $M_{\rm A}$ at scale $l$ is defined as the ratio of turbulent velocity in the volume to Alfv\'en speed: $M_{\rm A}=v_l/v_{\rm A}$. For Alfv\'enic turbulence, we have $v_l/v_{\rm A}=\delta B/\langle B\rangle$ so that $M_{\rm A}=\delta B/\langle B\rangle$.}. 

In the presence of a mean magnetic field, the integral of local $M_{\rm A}$ weighted by density $n$ can be replaced with its mean value $\overline {M_{\rm A}}$ averaged along the LOS, as a first order approximation. In this work, upper "$-$" symbol means LOS average, while $\langle...\rangle$ is averaged over a volume of interest. 

For convenience, we introduce $\overline M_{\rm A,\bot}$, which is the Alfv\'en Mach number corresponding to the motions perpendicular to the LOS, i.e.:
\begin{equation}
\begin{aligned}
   {\overline M_{\rm A,\bot}}&=(v_l\sqrt{4\pi{\overline\rho}})/(\langle B \rangle\sin{\overline\gamma})=v_l/({\overline v_A}\sin{\overline\gamma})\\
&={\overline M_{\rm A}}/\sin{\overline\gamma},
\end{aligned}
\label{Mperp}
\end{equation}
where ${\overline\rho}$ is mean gas mass density. The 3D turbulent velocity $v_l$ has been already incorporated in available observational methods (see \S~\ref{sec:dis}) of calculating
$\overline M_{\rm A,\bot}$. For instance, the Davis-Chandrasekhar-Fermi (DCF) method \citep{1951PhRv...81..890D,1953ApJ...118..113C} calculates $v_l$ from the emission line’s width \citep{2021ApJ...913...85H} \footnote{Note that the LOS turbulent velocity $v_{\rm los}$  calculated from the emission line, after correcting thermal speed and telescope beam effect \citep{1999ApJ...520..706C,2021ApJ...913...85H}, corresponds to the fluctuation at the injection scale which is isotropic. Consequently, the 3D turbulent velocity at injection scale $v_{\rm inj}$ can be obtained from $v_{\rm inj}=\sqrt{3}v_{\rm los}$ (see Appendix~\ref{appendix}). When turbulence cascades to a small scale, the fluctuation becomes anisotropic, i.e., most significant in the direction perpendicular to the magnetic field, as confirmed by numerical simulations \citep{HXL21}. The turbulent velocity $v_l$ at scale $l$ is $v_l=(l/L_{\rm inj})^{1/3}v_{\rm inj}$ for Kolmogorov-type turbulence, where $L_{\rm inj}$ is the injection scale. $v_l$ obtained in observation, therefore, contains the contribution not only from the LOS velocity.}. The projection, therefore, is applied only to the total magnetic field strength.

Combining Eqs.~\ref{eq.4} and \ref{eq:qu}, the polarization fraction can be written as:
\begin{equation}
\label{eq.pma}
    \begin{aligned}
p&=\frac{p_0}{1+\overline {M_{\rm A}}^2}\cdot\frac{\sin^2\langle\gamma\rangle(1-\frac{1}{2} \overline {M_{\rm A}}^2)}{1-p_0(1/3-\frac{\sin^2\langle\gamma\rangle(\overline {M_{\rm A}}^2-2)+2}{2(\overline {M_{\rm A}}^2+1)})}.\\
\end{aligned}
\end{equation}
Note here we write the Eq.~\ref{eq.4}'s integral of the product in the numerators' two terms and the denominator second term as a product of two integrals (one is $\int n dz$) as we disregard the correlation of fluctuations of density and magnetic field.  Consequently, the column density $\int n dz$ appears in both numerator and denominator and is cancelled off.  In reality, the observationally measured polarization angle and inclination angle are density weighted. However,
the main effect for polarization is expected from the variations of the magnetic field direction.\footnote{ Grain alignment by radiative torques \citep{Lazarian07,BG15} and related dust disruption \citep{2021ApJ...908...12L,2019ApJ...876...13H} can also vary for different LOS and affect polarization. These effects are expected for clouds with active star formation or for LOS with high optical depth. We disregard these effects within our model.}

\begin{figure}
	\includegraphics[width=1.0\linewidth]{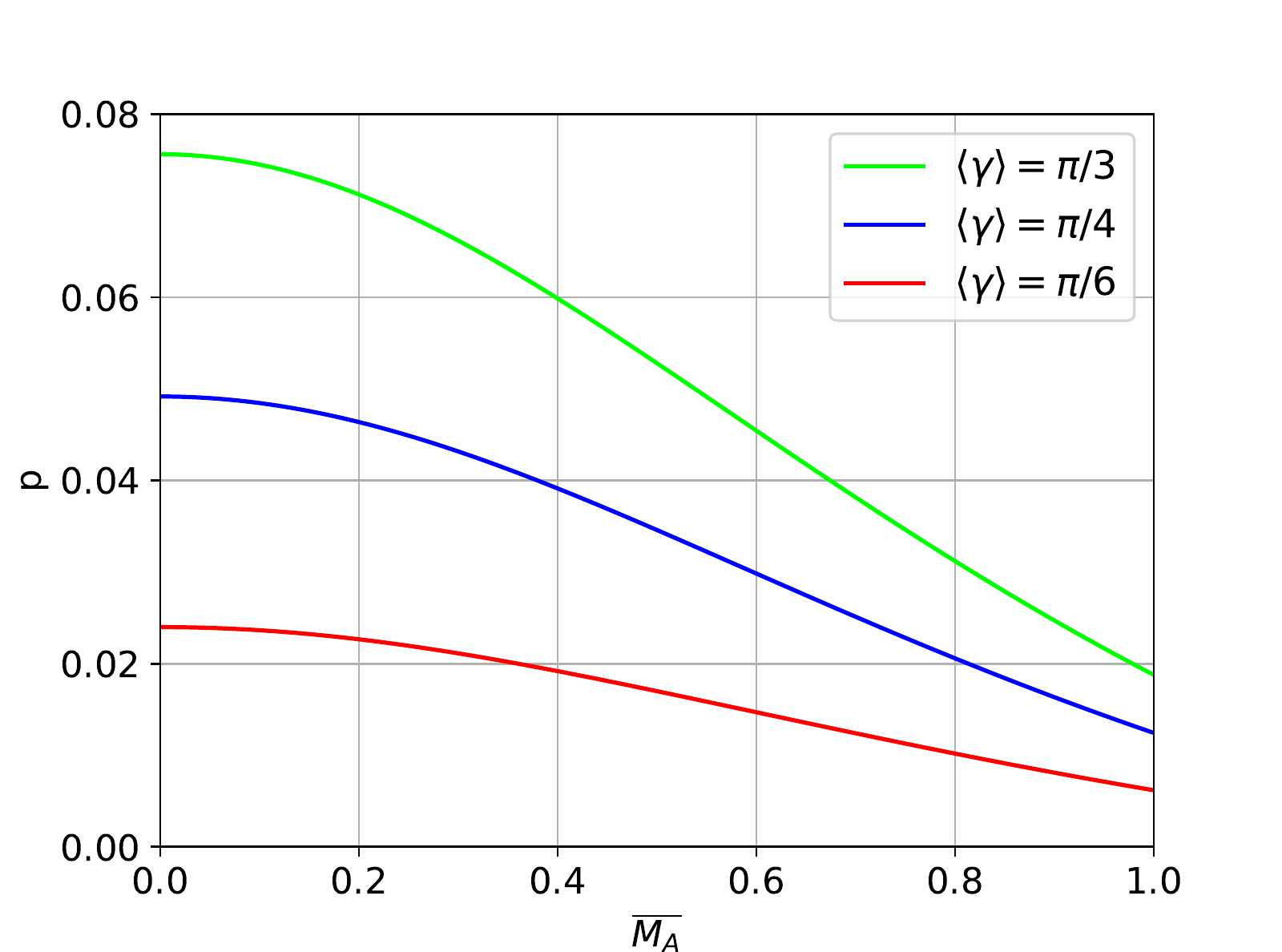}
    \caption{Analytical relation Eq.~\ref{eq.pma} of the polarization fraction $p$ and $\overline{M_{\rm A}}$. }
    \label{fig:p}
\end{figure}

As shown in Fig.~\ref{fig:p}, the  variations of the magnetic field direction along the LOS induce depolarization effects so that $p$ get its minimum value at large $\overline{M_{\rm A}}$. 
%
In observations, as $p$ is measured, the key problem in determining $\sin^2\langle\gamma\rangle$ is to get $p_0$ and $\overline {M_{\rm A}}$. \citet{2019MNRAS.485.3499C} showed that $p_0$ can be recovered approximately from: 
\begin{equation}
\label{eq:p0}
    p_0=\frac{3p_{\rm max}}{3+p_{\rm max}},
\end{equation}
where $p_{\rm max}$ is the maximum polarization fraction that can be obtained when the local inclination angle is 90$^\circ$. The discussion of uncertainties of $p_{\rm max}$ determination within our treatment is given in \S~\ref{sec:dis}.
If we know  $p_0$, we can express the distribution of total Mach number explicitly from the observed polarization fraction $p$:
\begin{equation}
\label{eq:ma}
\overline {M_{\rm A}}^2=\frac{p_0\sin^2\langle\gamma\rangle(1+p)-p(1+\frac{2}{3}p_0)}{\frac{1}{2}p_0\sin^2\langle\gamma\rangle(1+p)+p(1-\frac{1}{3}p_0)}.
\end{equation}
Note that the condition $\overline {M_{\rm A}}\ge0$ implicitly restricts the numerator to be non-negative. 

With the assumption of vanishing fluctuations ($\overline{M_{\rm A}}^2\approx0$), \cite{2019MNRAS.485.3499C} generalized Eq.~\ref{eq:ma} to every LOS to get local $\overline{\gamma}$ instead of the mean value $\langle\gamma\rangle$:
\begin{equation}
\label{eq:chen}
\sin^2\overline{\gamma}_{\rm Ch}=\frac{p(1+\frac{2}{3}p_0)}{p_0(1+p)},
\end{equation}
where the subscript "$\rm Ch$" is used to distinguish the expression in \citet{2019MNRAS.485.3499C} from our expression. Here we see an inconsistency in the treatment of the problem in \citet{2019MNRAS.485.3499C}. The condition $\overline{M_{\rm A}}^2\approx0$ cannot be satisfied for every LOS in observation. As we are interested in the realistic situation of $\overline{M_{\rm A}}$ being nonzero, by accounting for $\overline{M_{\rm A}}$, we address this inconsistency. Combining Eq.~\ref{Mperp} and Eq.~\ref{eq:ma}, the expression for local $\sin{\overline\gamma}$ is:
\begin{equation}
\label{eq.gammap}
\sin^2\overline{\gamma}=\frac{1}{\overline{M_{\rm A}}_{,\bot}^2}\cdot\frac{p_0\sin^2\langle\gamma\rangle(1+p)-p(1+\frac{2}{3}p_0)}{\frac{1}{2}p_0\sin^2\langle\gamma\rangle(1+p)+p(1-\frac{1}{3}p_0)}.
\end{equation}
Or alternatively, we have:
\begin{equation}
\label{eq:p}
    p=\frac{p_0}{1+\overline {M_{\rm A}}_{,\bot}^2\sin^2\overline{\gamma}}\cdot\frac{\sin^2\langle\gamma\rangle(1-\frac{1}{2} \overline {M_{\rm A}}_{,\bot}^2\sin^2\overline{\gamma})}{1-p_0(1/3-\frac{\sin^2\langle\gamma\rangle(\overline {M_{\rm A}}_{,\bot}^2\sin\overline{\gamma}^2-2)+2}{2(\overline {M_{\rm A}}_{,\bot}^2\sin^2\overline{\gamma}+1)})}.
\end{equation}

In the situation of the zeroth order approximation $\overline {M_{\rm A}}_{,\bot}^2\approx0$, the contribution from $\overline {M_{\rm A}}_{,\bot}^2\sin^2\overline{\gamma}$ vanished. Or in other situation that $\overline {M_{\rm A}}_{,\bot}^2$ is the leading term, the condition $\overline {M_{\rm A}}_{,\bot}^2\ll1$ also guarantees that $\overline {M_{\rm A}}_{,\bot}^2\sin^2\overline{\gamma}$ is negligible. Consequently, if one can find a LOS satisfying $\overline {M_{\rm A}}_{,\bot}^2\ll1$, Eq.~\ref{eq:p} reduces to:
\begin{equation}
\begin{aligned}
\label{eq.poff}
p_{\rm off}&=\frac{p_0\sin^2\langle\gamma\rangle_{\rm off}}{1-p_0(\sin^2\langle\gamma\rangle_{\rm off}-\frac{2}{3})},~~~\overline {M_{\rm A}}_{,\bot}^2\ll1,
\end{aligned}
\end{equation}
where $p_{\rm off}$ is the polarization fraction corresponding to $\overline {M_{\rm A}}_{,\bot}\ll1$. Equivalently, the mean inclination angle is:
\begin{equation}
\label{eq:siny}
\begin{aligned}
\sin^2\langle\gamma\rangle_{\rm off}&=\frac{p_{\rm off}(1+\frac{2}{3}p_0)}{p_0(1+p_{\rm off})},~~~\overline {M_{\rm A}}_{,\bot}^2\ll1,
\end{aligned}
\end{equation}
where the subscript "off" represents that the mean inclination angle is calculated with
the knowledge of polarization fraction and  $\overline{M_{\rm A}}_{,\bot}$ at a reference position. 
In this work, we explore the combination of Eqs.~\ref{eq:p0} and \ref{eq:siny} in obtaining three-dimensional magnetic field assuming $\overline{M_{\rm A}}_{,\bot}^2$ is the leading term:
\begin{equation}
\label{eq.17}
    \sin^2\langle\gamma\rangle_{\rm off}=\frac{p_{\rm off}(1+p_{\rm max})}{p_{\rm max}(1+p_{\rm off})},~~~\overline {M_{\rm A}}_{,\bot}^2\ll1.
\end{equation}

Also, the total mean Alfv\'en Mach number can be naturally accessed via $\langle\overline {M_{\rm A}}_{,\bot}\rangle\sin\langle{\gamma}\rangle_{\rm off}=\langle\overline {M_{\rm A}}\rangle$. Note that Eq.~\ref{eq:p0} assumes that local inclination angle can achieve 90$^\circ$, which, however, might not be the case in observation. We, therefore, generalize $p_{\max}$ to the maximum value of observed polarization fraction. Although this generalization introduces uncertainty to the estimation of $\sin^2\langle\gamma\rangle_{\rm off}$, we numerically find it is insignificant (see \S~\ref{sec:result}). 

Moreover, Eq.~\ref{eq.17} requires the information of $\overline{M_{\rm A}}_{,\bot}$ to estimate $\langle\gamma\rangle_{\rm off}$. However,
observations of dust polarization allow to measure the magnetic field's variation $\delta \phi$ perpendicular to the LOS.
In the case that the polarization's integration length scale along the LOS does not exceed the turbulent injection scale, one can introduce the relation ${\overline M_{\rm A,\bot}}\approx{\overline {\delta \phi}}$ \citep{2008ApJ...679..537F,Lazarian18}. This approximation can be easily understood based on the fact that fluctuations are more significant for a weak magnetic field (i.e., large ${\overline M_{\rm A,\bot}}$). It is approximately true for molecular clouds and it is implicitly employed in the traditional treatment of DCF method to finding the strength of magnetic field. Thus with polarization measurement alone, one can still estimate $\langle\gamma\rangle_{\rm off}$ from:
\begin{equation}
\label{eq.18}
    \sin^2\langle\gamma\rangle_{\rm off}=\frac{p_{\rm off}(1+p_{\rm max})}{p_{\rm max}(1+p_{\rm off})},~~~{\overline {\delta \phi}}^2\ll1,
\end{equation}
where $\delta \phi$ that we associate with $M_{A,\bot}$ should be determined statistically. Therefore, we deal with a statistically averaged quantities, similar to what is done in the DCF method.

For the simplicity of test, we calculate the distribution of $\overline{M_{\rm A}}_{,\bot}$ from numerical simulations directly. To implement it in observation, additional approaches of measuring $\overline{M_{\rm A}}_{,\bot}$ or $\delta \phi$ are required. We list several possible solutions in \S~\ref{sec:dis} and one observational implementation in \citet{2022arXiv221011023H}.

\subsection{Perturbation expansion}
As suggested by Eq.~\ref{eq:qu}, the magnetic fluctuation magnifies further depolarization. Here we consider a more general form of perturbation expansion to investigate its significance. We introduce $\lambda$ as a dimensionless parameter that can take on values ranging continuously from 0 (no fluctuation) to 1 (the full fluctuation):
\begin{equation}
    \pmb{B}(x,y,z)=\langle \pmb{B}\rangle+\lambda\delta \pmb{B}(x,y,z).
\end{equation}
Consequently, the $Q$ and $U$ in Eq.~\ref{eq:qu} becomes:
\begin{equation}
\begin{aligned}
Q&=\int p_0n\frac{\cos(2\langle\psi\rangle)[\sin^2\langle\gamma\rangle+ \frac{\lambda^2}{2}M_{\rm A}^2\cos^2\langle\gamma\rangle-\frac{\lambda^2}{2} M_{\rm A}^2]}{1+\lambda^2M_{\rm A}^2}dz,\\
U&=\int p_0n\frac{\sin(2\langle\psi\rangle)[\sin^2\langle\gamma\rangle+\frac{\lambda^2}{2} M_{\rm A}^2\cos^2\langle\gamma\rangle-\frac{\lambda^2}{2} M_{\rm A}^2]}{1+\lambda^2M_{\rm A}^2}dz.\\ 
\end{aligned}
\end{equation}
In the case that the fluctuation is sufficiently weak, $Q$ and $U$ can be written as a power series in $\lambda$:
\begin{equation}
\begin{aligned}
Q&\approx\sum_{n=0}^{2}\lambda^n\frac{1}{n!}\frac{d^nQ}{d\lambda^n}\rvert_{\lambda=0}=\int p_0n\cos(2\langle\psi\rangle)\sin^2\langle\gamma\rangle(1-3\lambda^2M_{\rm A}^2)dz,\\
U&\approx\sum_{n=0}^{2}\lambda^n\frac{1}{n!}\frac{d^nU}{d\lambda^n}\rvert_{\lambda=0}=\int p_0n\sin(2\langle\psi\rangle)\sin^2\langle\gamma\rangle(1-3\lambda^2M_{\rm A}^2)dz,\\ 
\end{aligned}
\end{equation}
here we expand the $Q$ and $U$ only to the second-order. We notice that the first-order expansion vanishes because of $\frac{dQ}{d\lambda}\rvert_{\lambda=0}=0$, $\frac{dU}{d\lambda}\rvert_{\lambda=0}=0$. It suggests that the depolarization contributed by the fluctuation in magnetic field is a second-order quantity. The primary source of depolarization is the inclination angle's fluctuation $3\sin^2\langle\gamma\rangle M_{\rm A}^2$. 

\subsection{Sub-region sampling}
Eq.~\ref{eq.17} could reveal the mean inclination angle for a given cloud under the assumption that $p_0$ is constant across the entire cloud and dust grains' properties are homogeneous. We denote this method as Polarization Fraction Analysis (PFA).

The accuracy of the PFA mainly depends on (i) the presence of a mean magnetic field; (ii) the existence of a reference position with $\overline {M_{\rm A}}_{,\bot}^2\ll1$ assuming $\overline{M_{\rm A}}_{,\bot}$ is the leading factor in Eq.~\ref{eq:p}; (iii) the samples within a region are sufficient so that our assumption of perpendicular magnetic field fluctuations is valid; and (iv) the whether the maximum value $p_{\rm max}$ of observed polarization fraction corresponds to the case that the local inclination angle is 90$^\circ$. We will numerically show in \S~\ref{sec:result} that the underestimation of $p_{\rm max}$ has insignificant effect.

The four conditions, more or less, are related to the number of samples within a region. Therefore, it is not necessary to choose the full cloud as the object for the application. Once the four conditions are satisfied for a sub-region within the cloud, the PFA is applicable. We denote this zoom-in procedure as sub-region sampling.

\begin{table}
	\centering
	\label{tab1}
	\begin{tabular}{| c | c | c | c | c |}
		\hline
		Model & $M_s$ & $M_{\rm A}$ & Resolution & $\beta$ \\	\hline\hline
		A0 & 5.38 & 0.41 & $792^3$ & 0.01 \\
		A1 & 5.40 & 0.61 & $792^3$ & 0.03 \\
		A2 & 5.23 & 0.95 & $792^3$ & 0.07 \\
	    A3 & 5.12 & 1.13 & $792^3$ & 0.10 \\\hline
	\end{tabular}
	\caption{Description of MHD simulations. The compressibility of turbulence is characterized by $\beta=2(\frac{M_{\rm A}}{M_s})^2$.}
\end{table}

\section{Numerical method}
\label{sec:data}
The numerical simulations used in this work are generated through ZEUS-MP/3D code \citep{2006ApJS..165..188H}. We simulate an isothermal cloud in the Eulerian frame by solving the ideal MHD equations with periodic boundary conditions. The cloud is initiated with uniform density field $\langle\rho\rangle$ and magnetic field $\langle\pmb{B}\rangle$ along the x-axis, which is perpendicular to the LOS. 

We are considering pure turbulence cases without self-gravity. Kinetic energy is solenoidally injected at wavenumber $\sim2$ to produce a Kolmogorov spectrum. The solenoidal driving mechanism can also generate a compressive component. We continuously drive turbulence and dump the data until the turbulence gets fully developed at one sound crossing time. The simulation is grid into 792$^3$ cells, and turbulence gets numerically dissipated at scales $\approx$ 10 - 20 cells. Turbulence induces  magnetic field fluctuation $\delta\boldsymbol{B}$ and density fluctuation $\delta\rho$ accordingly

Simulation of MHD turbulence is scale-free. Its properties are characterized by the sonic Mach number $M_s=v_{\rm inj}/c_{s}$ and Alfv\'{e}nic Mach number $M_{\rm A}=v_{\rm inj}/v_{\rm A}$, where $v_{\rm inj}$ is the velocity fluctuation at injection scale. The sound speed $c_s\approx0.192$ in the code unit is fixed due to the isothermal equation of state. To simulate different ISM conditions, we change the initial uniform magnetic field and density field, as well as the injected kinetic energy to achieve various $M_{\rm A}$ and $M_s$ values. In this work, we refer to the simulations in Tab.~\ref{tab1} by their model name or key parameters. Similar simulations have been used in \citet{2020ApJ...905..129H}.

Synthetic dust emission is then calculated from Eq.~\ref{eq.dust} by extracting the necessary information from the MHD simulation. We assume a constant intrinsic polarization fraction $p_0=0.1$. The mean inclination angle $\langle\gamma\rangle$ of the simulation is calculated from:
\begin{equation}
\begin{aligned}
    \langle\gamma\rangle&=\cos^{-1}(\frac{\langle B_z\rangle}{\langle B\rangle}).
\end{aligned}
\end{equation}
Note here $\langle...\rangle$ means averaging over all cells. We rotate the simulation box to achieve different inclination angles.

In particular, $M_{\rm A}^{\rm 3D}$ at a cell and its POS projection $M_{\rm A,\bot}$ are approximated by:
\begin{equation}
\begin{aligned}
M_{\rm A}^{\rm 3D}&=(|\pmb{B}-\langle \pmb{B}\rangle|)/\langle B\rangle,\\
\gamma_{\rm 3D}& =\cos^{-1}(\frac{|B_z|}{B}),\\
M_{\rm A,\bot}&=\frac{M_{\rm A}^{\rm 3D}}{\sin\gamma_{\rm 3D}},
\end{aligned}
\end{equation}
where $\gamma_{\rm 3D}$ is the local inclination angle at a cell.
Averaging $M_{\rm A,\bot}$ along each LOS gives $\overline{M_{\rm A}}_{,\bot}$ accordingly.

We compare the global inclination angle estimated by our approach with the one proposed by \cite{2019MNRAS.485.3499C}. We denote the mean inclination angle inferred from Eq.~\ref{eq.17} as:
\begin{equation}
\begin{aligned}
\langle\gamma\rangle_{\rm off}&=\sqrt{\sin^{-1}[\frac{p_{\rm off}(1+p_{\rm max})}{p_{\rm max}(1+p_{\rm off})}]},~~~\overline {M_{\rm A}}_{,\bot}^2\ll1,
\end{aligned}
\end{equation}
and the one calculated from \citet{2019MNRAS.485.3499C} as:
\begin{equation}
\begin{aligned}
\overline{\gamma}_{\rm Ch19}&=\sqrt{\sin^{-1}[\frac{p(1+\frac{2}{3}p_0)}{p_0(1+p)}]},\\
\langle\gamma\rangle_{\rm Ch19}&=\tan^{-1}(\frac{\langle\sin\overline{\gamma}_{\rm Ch19}\rangle}{\langle\cos\overline{\gamma}_{\rm Ch19}\rangle}).
\end{aligned}
\end{equation}

\begin{figure}
	\includegraphics[width=1.0\linewidth]{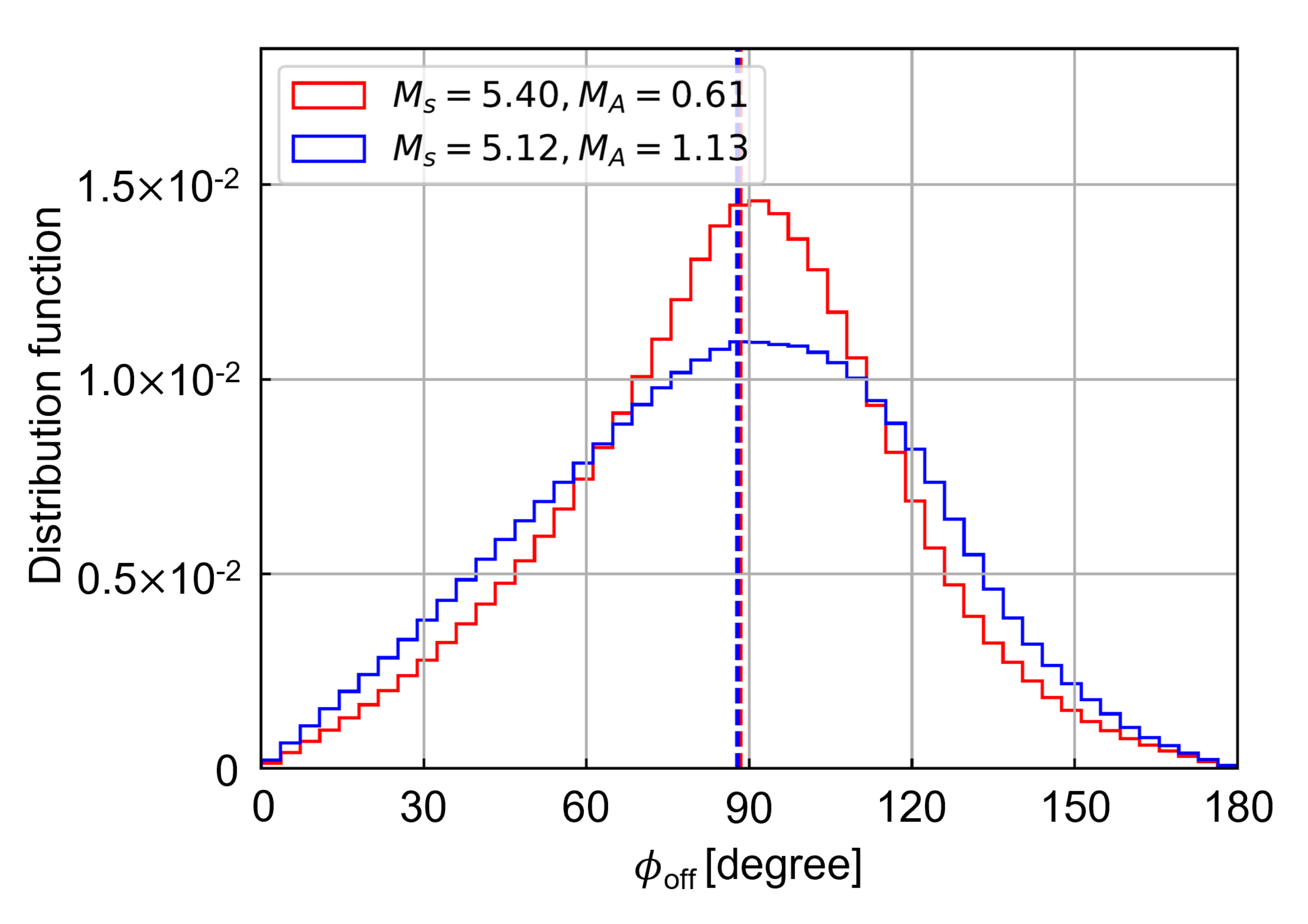}
    \caption{Histogram of the relative angle $\phi_{\rm off}$ between the magnetic field fluctuation $\delta\pmb{B}$ and mean magnetic field $\langle\pmb{B}\rangle$. Dashed line indicates the median value. Mean inclination angle in the simulations is $90^\circ$.}
    \label{fig:hist}
\end{figure}

\begin{figure}
	\includegraphics[width=1.0\linewidth]{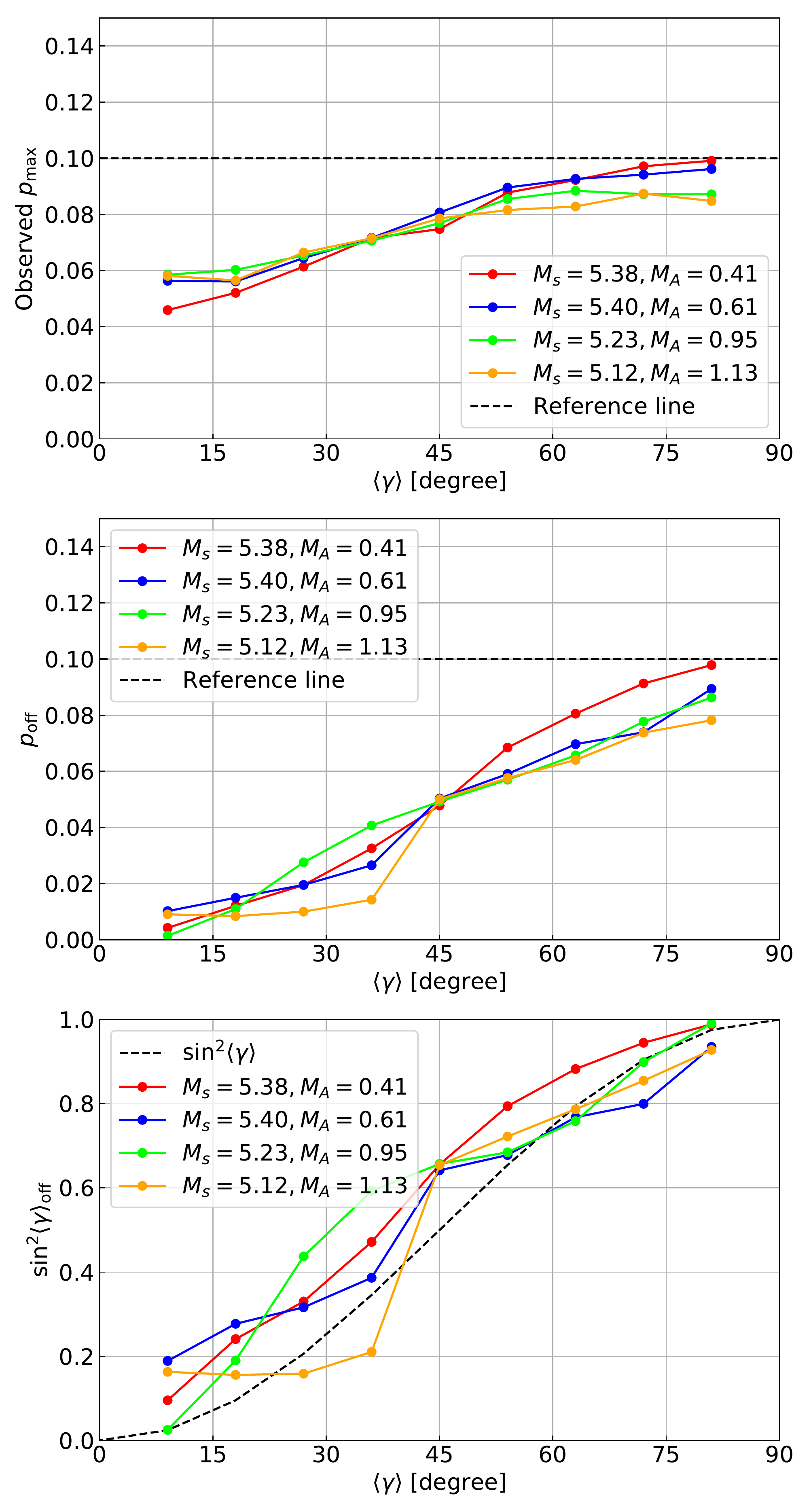}
    \caption{The observed $p_{\rm max}$ (top), $p_{\rm off}$ (middle), and estimated $\sin^2\langle\gamma\rangle_{\rm off}$ (bottom) as a function of the actual mean inclination angle $\langle\gamma\rangle$. The reference lines in the top two panels represent the intrinsic polarization fraction in simulations.}
    \label{fig:pmax}
\end{figure}

\begin{figure}
	\includegraphics[width=1.0\linewidth]{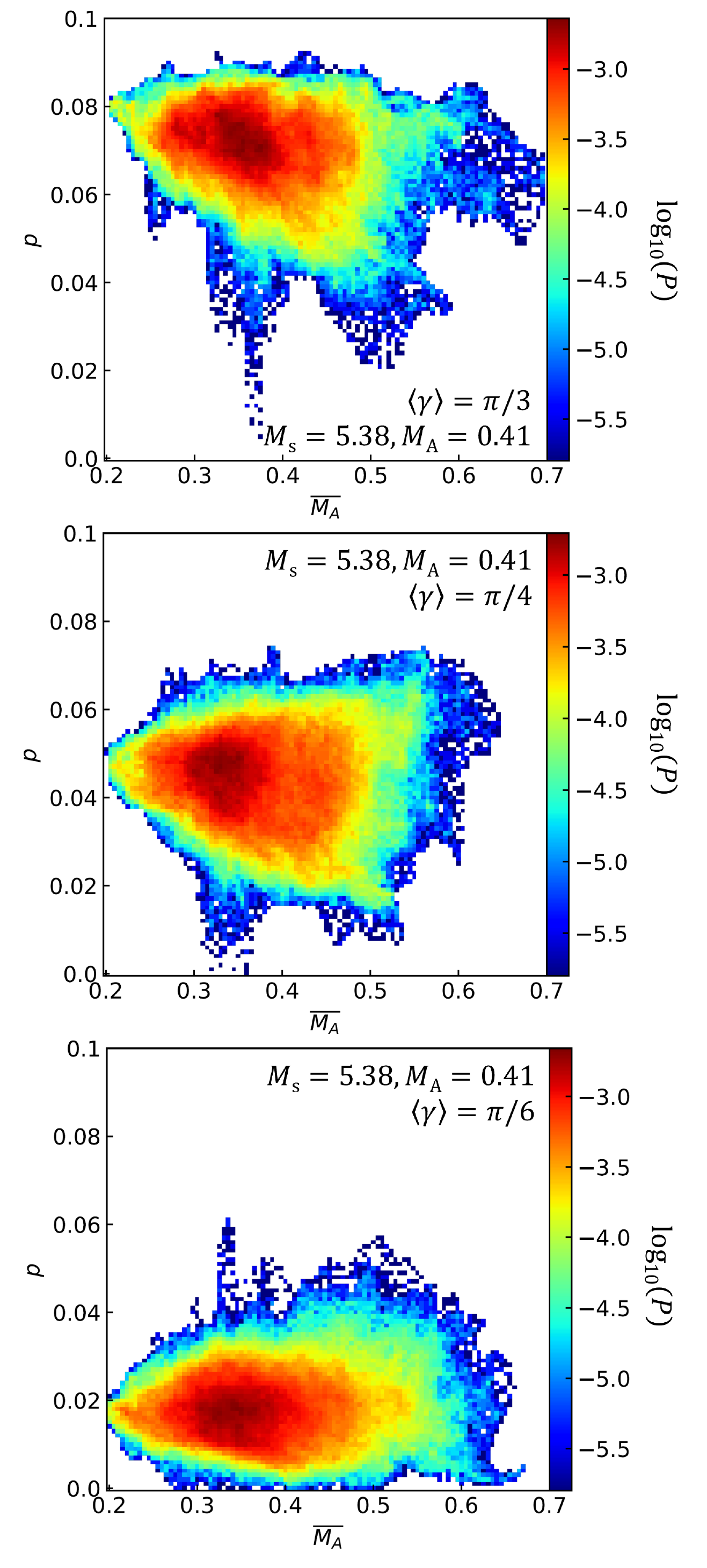}
    \caption{2D histogram of polarization fraction $p$ and averaged total Alfv\'en Mach number $\overline{M_{\rm A}}$ in the conditions of various mean inclination angle $\langle\gamma\rangle$. $P$ denotes the percent of sampling points. }
    \label{fig:pma}
\end{figure}

\begin{figure*}
	\includegraphics[width=1.0\linewidth]{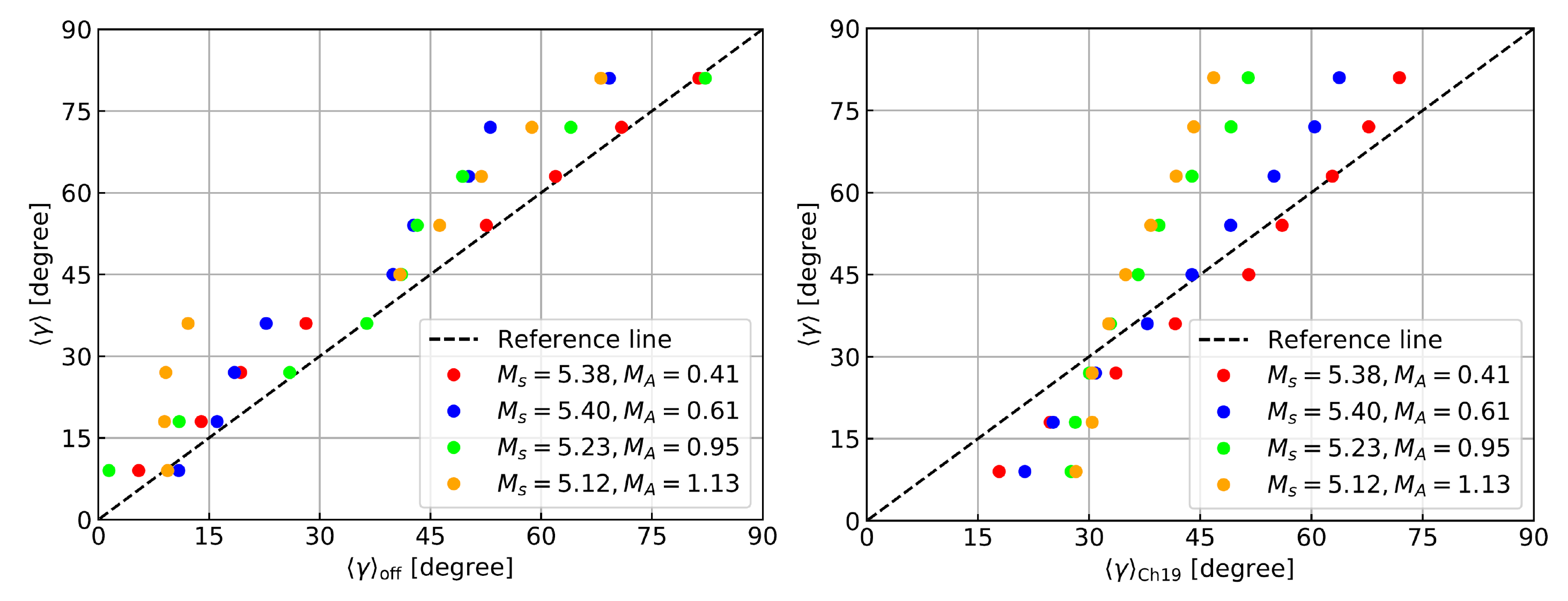}
    \caption{Comparison of the mean inclination angle $\langle\gamma\rangle_{\rm off}$/ $\langle\gamma\rangle_{\rm Ch19}$ (left/right) with the real inclination angle $\langle\gamma\rangle$ of the simulation.  $\langle\gamma\rangle_{\rm off}$ is derived in this work, while $\langle\gamma\rangle_{\rm Ch19}$ was proposed by Chen et al. (2019). }
    \label{fig:gamma}
\end{figure*}

\begin{figure*}
	\includegraphics[width=0.9\linewidth]{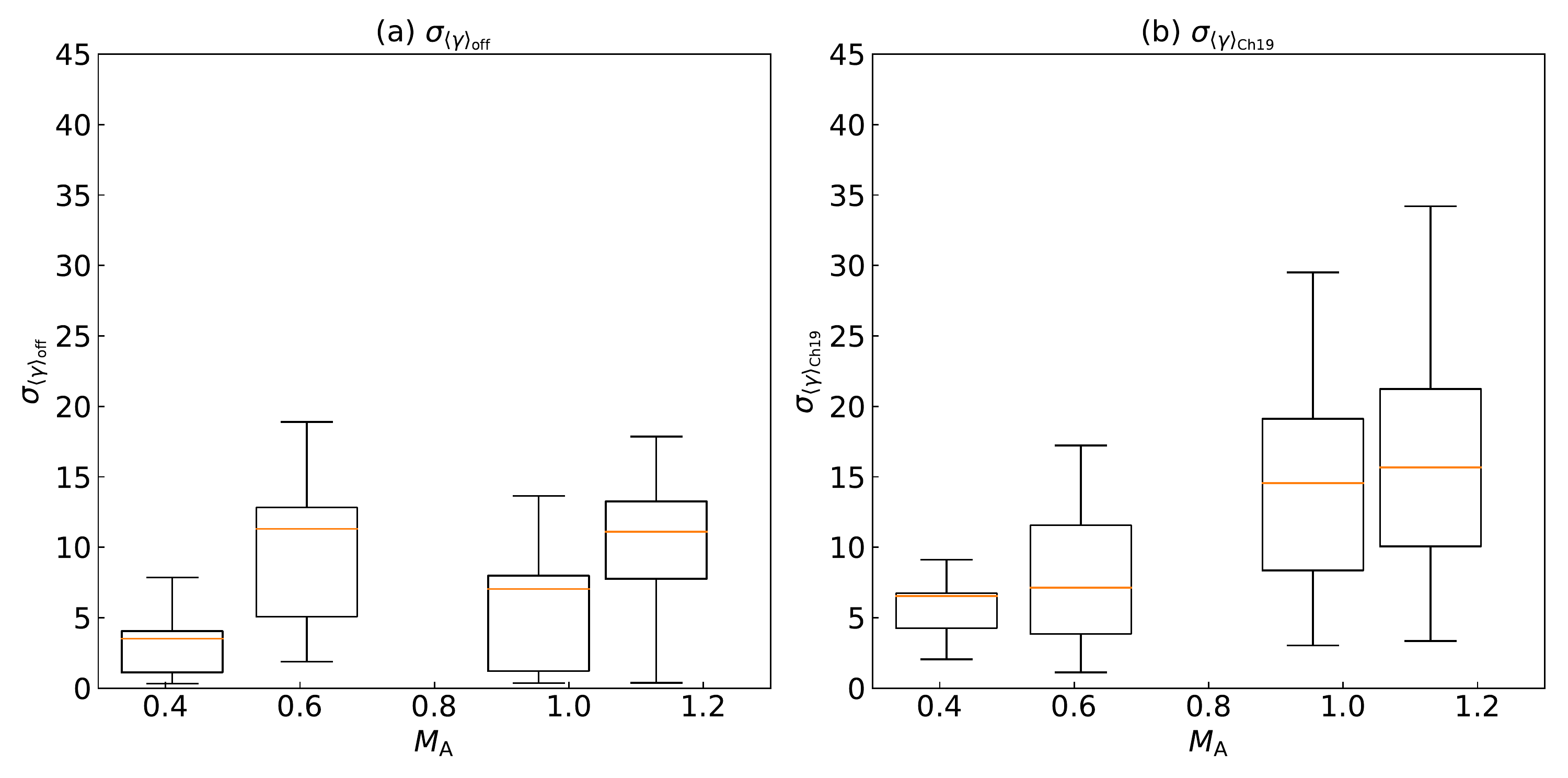}
    \caption{Deviation of estimated inclination angle and actual inclination angle. Upper and lower black lines represent the deviation's maximum and minimum, respectively. Box gives ranges of the first (lower) and third quartiles (upper) and orange line represent the median value. \textbf{Panel a:}    $\sigma_{\langle\gamma\rangle_{\rm off}}$ in degrees represents the absolute difference between $\langle\gamma\rangle_{\rm off}$ and $\langle\gamma\rangle$. \textbf{Panel b:} $\sigma_{\langle\gamma\rangle_{\rm Ch19}}$ in degrees is for the absolute difference of $\langle\gamma\rangle_{\rm Ch19}$ and $\langle\gamma\rangle$.
    }
    \label{fig:sigma}
\end{figure*}

The relative orientation between the measured inclination angle and real inclination angle of the simulation is measured with the Alignment Measure (AM; \citealt{GL17}), defined as:
\begin{equation}
        \begin{aligned}
           {\rm AM}  =  2(\langle\cos^2\theta_r\rangle-\frac{1}{2}),
        \end{aligned}
\end{equation}
where $\theta_r$ is the relative angle between two vectors. AM is an averaged quantity, and its value is in the range of [-1, 1]. AM = 1 indicates that two sets of vectors are parallel, and AM = -1 denotes that the two are orthogonal.

\begin{figure*}
	\includegraphics[width=1.0\linewidth]{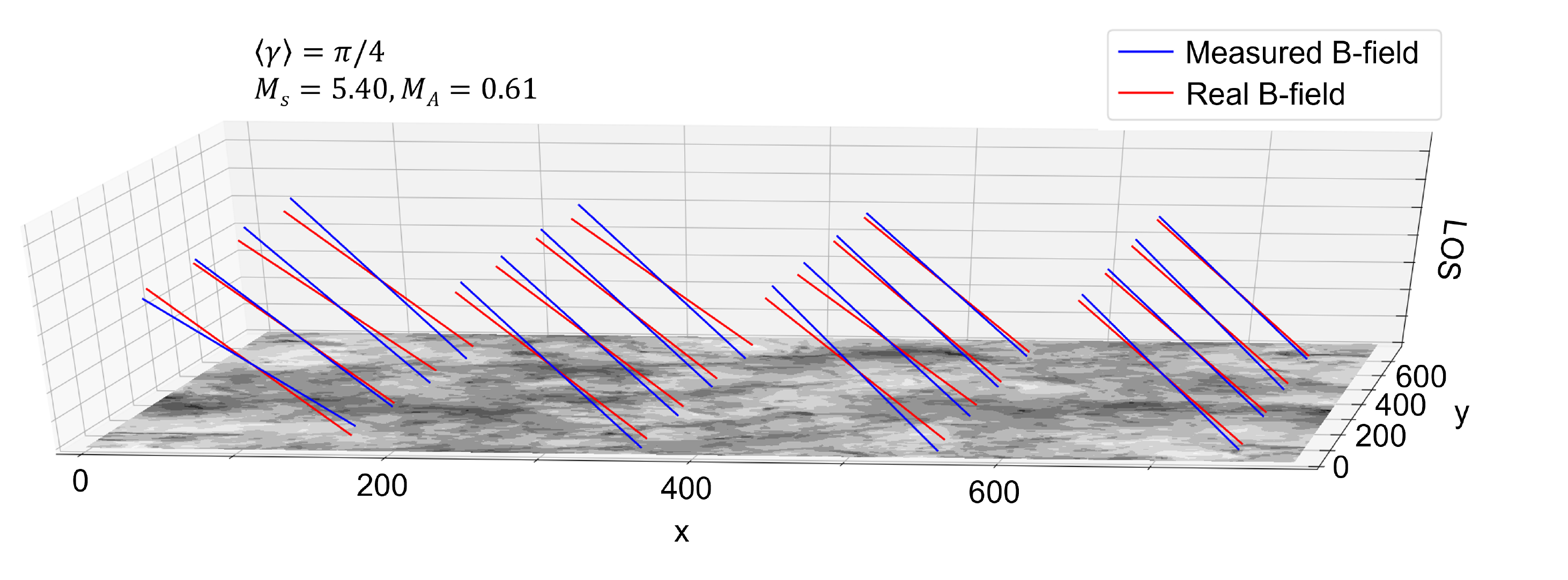}
    \caption{An example of the inclination angles measured for sixteen sub-regions with size $198\times198$ cell$^2$.  Each magnetic field vector is constructed by the POS magnetic field's position angle (i.e., $\psi+\pi/2$) inferred from Stokes parameters and the inclination angle of either measured $\langle\gamma\rangle_{\rm off}$ (blue) or actual $\langle\gamma\rangle_{\rm sub}$ (red). Note that the obtained magnetic field is the projection along the LOS. The third axis of LOS is for 3D visualization purposes having no distance information here. The total intensity map $I$ is placed on the POS, i.e., the $x-y$ plane. The axis's length ratio is 1:1:1 when plotting the vectors.}
    \label{fig:3D}
\end{figure*}

\begin{figure*}
	\includegraphics[width=1.0\linewidth]{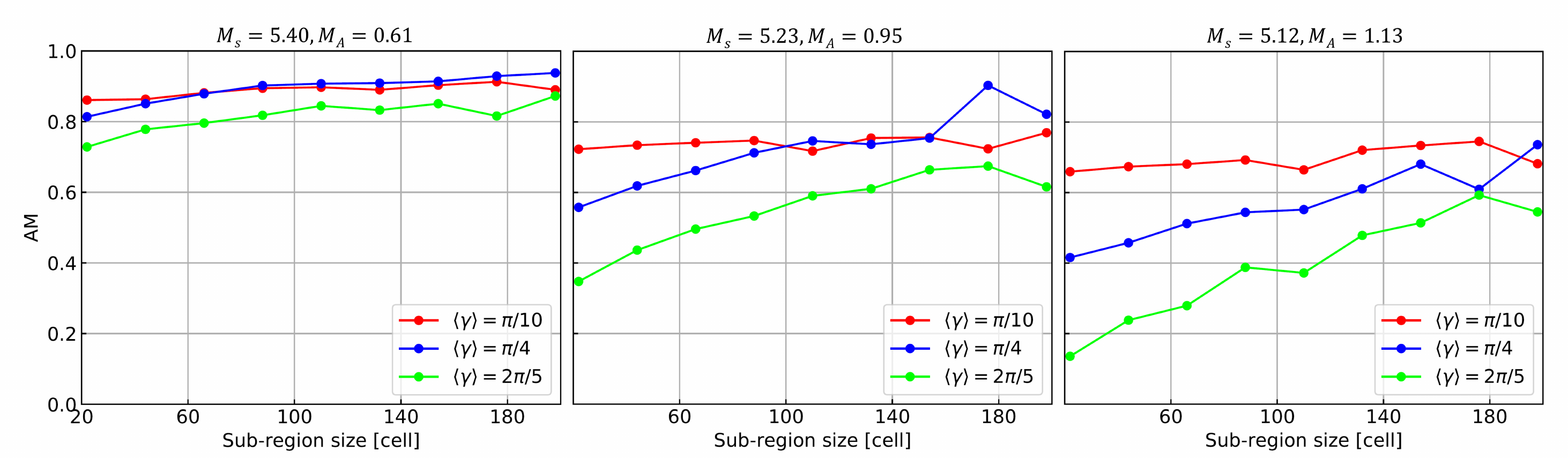}
    \caption{The AM of $\langle\gamma\rangle_{\rm off}$ and $\langle\gamma\rangle_{\rm sub}$ as a function of the sub-region's size. $\langle\gamma\rangle$ denotes the global mean inclination angle of the full simulation box.}
    \label{fig:sb}
\end{figure*}

\section{Results}
\label{sec:result}
\subsection{The relative angle of mean magnetic field and fluctuations}
Fig.~\ref{fig:hist} presents the histogram of the relative angle $\phi_{\rm off}$ between the magnetic field fluctuation $\delta\pmb{B}$ and mean magnetic field $\langle\pmb{B}\rangle$. The adopted simulations consist of compressible turbulence rather than only incompressible turbulence. However, we can see that for both sub-Alv\'enic and super-Alv\'enic cases, the histogram is close to a nearly symmetric distribution with a median value concentrated on $90^\circ$ around. The super-Alv\'enic case has a larger dispersion due to relatively stronger turbulence.

This median value of $\phi_{\rm off}\approx90^\circ$ is crucial for our assumption that the magnetic field's fluctuation preferentially appears in the mean field's perpendicular direction. This assumption is also valid in compressible turbulence.

\subsection{Effect of $p_{\rm max}$'s underestimation}
Eq.~\ref{eq:p0} is crucial in deriving the inclination angle using Eq.~\ref{eq.17}. It requires the value of $p_{\rm max}$, which corresponds to the case of local inclination angle $\sim90^\circ$, to estimate the intrinsic polarization fraction $p_0$. In a real scenario, this might not always be achieved. When the mean inclination angle is small, it is more difficult to locally achieve $\sim90^\circ$ . The only available information in observation is the maximum value of observed $p$, which does not necessarily correspond to the case that local inclination angle $\sim90^\circ$. Therefore, for practical application, we can only generalize Eq.~\ref{eq:p0} to the maximum value of observed $p$ and we denote this value as the observed $p_{\rm max}$. This generalization might underestimate $p_0$ and introduce uncertainty to the estimated mean inclination angle. 

In Fig.~\ref{fig:pmax}, we study the effect of $p_{\rm max}$'s underestimation in calculating $\sin^2\langle\gamma\rangle_{\rm off}$ assuming homogeneous dust properties. The maximum intrinsic polarization fraction in simulations is $\sim0.1$. However, we can see that the observed $p_{\rm max}$ achieves this value $\sim0.1$ only when the mean inclination angle $\langle\gamma\rangle$ is larger than $\sim60^\circ$. When $\langle\gamma\rangle<60^\circ$, the observed $p_{\rm max}$ rapidly decreases to $\sim0.05$, because local inclination angle cannot achieve $\sim90^\circ$. However, we find the decreasing trend of observed $p_{\rm max}$ when $\langle\gamma\rangle$ gets smaller is independent of $M_{\rm A}$, which characterizes the significance of magnetic field strength's fluctuation, i.e., strength of the fluctuations relative to the strength of the mean field, across the cloud. This suggests that the major depolarization agent is the inclination angle rather than magnetic field strength's fluctuation.

In addition to the observed $p_{\rm max}$, the value of $p_{\rm off}$ is also required to calculate $\sin^2\langle\gamma\rangle_{\rm off}$ (see Eq.~\ref{eq.17}). Here we obtain $p_{\rm off}$ from the polarization fraction corresponding to the minimum of $\overline {M_{\rm A}}_{,\bot}$. Due to statistically sufficient samples in the simulation, this choice satisfies the condition that $\overline {M_{\rm A}}_{,\bot}^2\ll1$. As shown in Fig.\ref{fig:pmax}, $p_{\rm off}$ rapidly decreases in the case of small $\langle\gamma\rangle$. $p_{\rm off}$ is already close to $\sim0$ when $\langle\gamma\rangle<10^\circ$. Similar to the case of observed $p_{\rm max}$, $p_{\rm off}$ has little dependence on $M_{\rm A}$.

Moreover, we find the calculated value of $\sin^2\langle\gamma\rangle_{\rm off}$ well follows the reference line of $\sin^2\langle\gamma\rangle$ when $\langle\gamma\rangle>45^\circ$. $\sin^2\langle\gamma\rangle_{\rm off}$ deviates more for small $\langle\gamma\rangle$ due to the underestimation of $p_{\rm max}$. We will quantify this uncertainty in the following. 

\subsection{Inclination angle as the major depolarization agent}
In general, in addition to the mean inclination angle and its fluctuation, magnetic field strength's fluctuation also contributes to the depolarization effect. However, as we see in Fig.~\ref{fig:pmax}, the inclination angle dominates the depolarization, while magnetic field strength's fluctuation gives an insignificant contribution. Moreover, the supersonic simulations of compressible MHD turbulence used in Fig.~\ref{fig:pmax} consist of significant density fluctuations. The observed $p_{\rm max}$, however, still achieves $\sim0.1$ when $\langle\gamma\rangle>75^\circ$. It suggests that density fluctuation contributes little to depolarization.

Fig.~\ref{fig:pma} presents the 2D histograms of polarization fraction $p$ and averaged total Alfv\'en Mach number $\overline{M_{\rm A}}$ along the LOS using the simulation A0. The histogram concentrates in a narrow range of $p$ when $\overline{M_{\rm A}}$ is relatively small, i.e., approximately $<0.4$. The histogram spreads to a wider range of $p$ when $\overline{M_{\rm A}}>0.4$. This more dispersed correlation is mainly caused by the inclination angle's fluctuation instead of magnetic field strength's fluctuation. When $\overline{M_{\rm A}}$ is large, significant fluctuations appear in both inclination angle and magnetic field strength. Because the inclination angle is the major agent for depolarization, its fluctuation, in this case, causes a rapid variation of $p$. Also, due to this effect, the observed $p_{\rm max}$ is more likely to appear in a position with relatively large $\overline{M_{\rm A}}$. This position locally achieves a large inclination angle so that the depolarization effect is relatively weak. 

 

\subsection{Comparison with Chen et al. (2019)}
Fig.~\ref{fig:gamma} presents the comparison of the full simulation cube's mean inclination angle obtained from Eq.~\ref{eq.17} with the one calculated from \cite{2019MNRAS.485.3499C}'s method. For $\langle\gamma\rangle_{\rm off}$ calculated through our method, generally, it is well compatible with the actual inclination angle $\langle\gamma\rangle$ of the simulation, although $\langle\gamma\rangle_{\rm off}$ gives slightly underestimated values. This underestimation might come from two reasons: (i) the underestimation of $p_{\rm max}$ as we discussed above; (ii) density fluctuation in compressible turbulence. Eq.~\ref{eq.17} is derived from the condition of incompressible turbulence, which contains no density fluctuation. It is possible that density fluctuation introduces uncertainties, although not significant.

As for \cite{2019MNRAS.485.3499C}'s method, its estimation agrees with $\langle\gamma\rangle$ better in strong magnetic field cases, i.e., sub-Alfv\'enic $M_{\rm A} = 0.41$ and $0.61$. $\langle\gamma\rangle_{\rm Ch19}$, however, significantly deviates from $\langle\gamma\rangle$ when $M_{\rm A} > 0.61$. This is caused by significant fluctuations in weakly magnetized turbulence, which breaks \cite{2019MNRAS.485.3499C}'s assumption that the fluctuations are negligible.

Fig.~\ref{fig:sigma} shows the deviation of the estimated inclination angle and actual angle. We calculate the absolute difference between $\langle\gamma\rangle_{\rm off}$ (or $\langle\gamma\rangle_{\rm Ch19}$) and $\langle\gamma\rangle$. The calculation is performed over all data points shown in Fig.~\ref{fig:gamma} and we denotes the difference as $\sigma_{\langle\gamma\rangle_{\rm off}}$ (or $\sigma_{\langle\gamma\rangle_{\rm Ch19}}$). Generally we see that the median value of $\sigma_{\langle\gamma\rangle_{\rm Ch19}}$ monotonically increases when $M_{\rm A}$ increases. It increases from $\approx6^\circ$ ($M_{\rm A}=0.41$) to $\approx15^\circ$ ($M_{\rm A}=1.13$). The trend of $\sigma_{\langle\gamma\rangle_{\rm off}}$'s median value is more complicated. It is similar to $\sigma_{\langle\gamma\rangle_{\rm Ch19}}$ in sub-Alfv\'enic case $M_{\rm A}<0.61$. In trans- and super-Alfv\'enic cases, $\sigma_{\langle\gamma\rangle_{\rm off}}$'s median stays in $10^\circ$ around. In addition to median value, the maximum $\sigma_{\langle\gamma\rangle_{\rm Ch19}}$ significantly increases to $\sim35^\circ$ in trans- and super-Alfv\'enic conditions, which comes from $\langle\gamma\rangle_{\rm Ch19}$'s underestimation in large $\langle\gamma\rangle$ cases (see Fig.~\ref{fig:gamma}). In general, $\sigma_{\langle\gamma\rangle_{\rm off}}$ ranges from 0 to $\sim20^\circ$ with a median value $\le10^\circ$, while $\sigma_{\langle\gamma\rangle_{\rm Ch19}}$ is in the range of 0 to $\sim35^\circ$. 


\subsection{Sub-region sampling}
As discussed above, our method mainly depends on three conditions:
(i) the existence of a mean magnetic field; (ii) the existence of a reference position with $\overline {M_{\rm A}}_{,\bot}^2\ll1$; (iii) the number of the sample within a region is sufficient so that perpendicular magnetic field fluctuations dominate. Thus, it is not necessary to perform the calculation to the full cloud or simulation. This method can be generalized to sub-regions satisfied with the conditions. In this section, we test the relation of $\langle\gamma\rangle_{\rm off}$'s accuracy and the sub-regions size. 
\begin{figure}
\centering
	\includegraphics[width=1.15\linewidth]{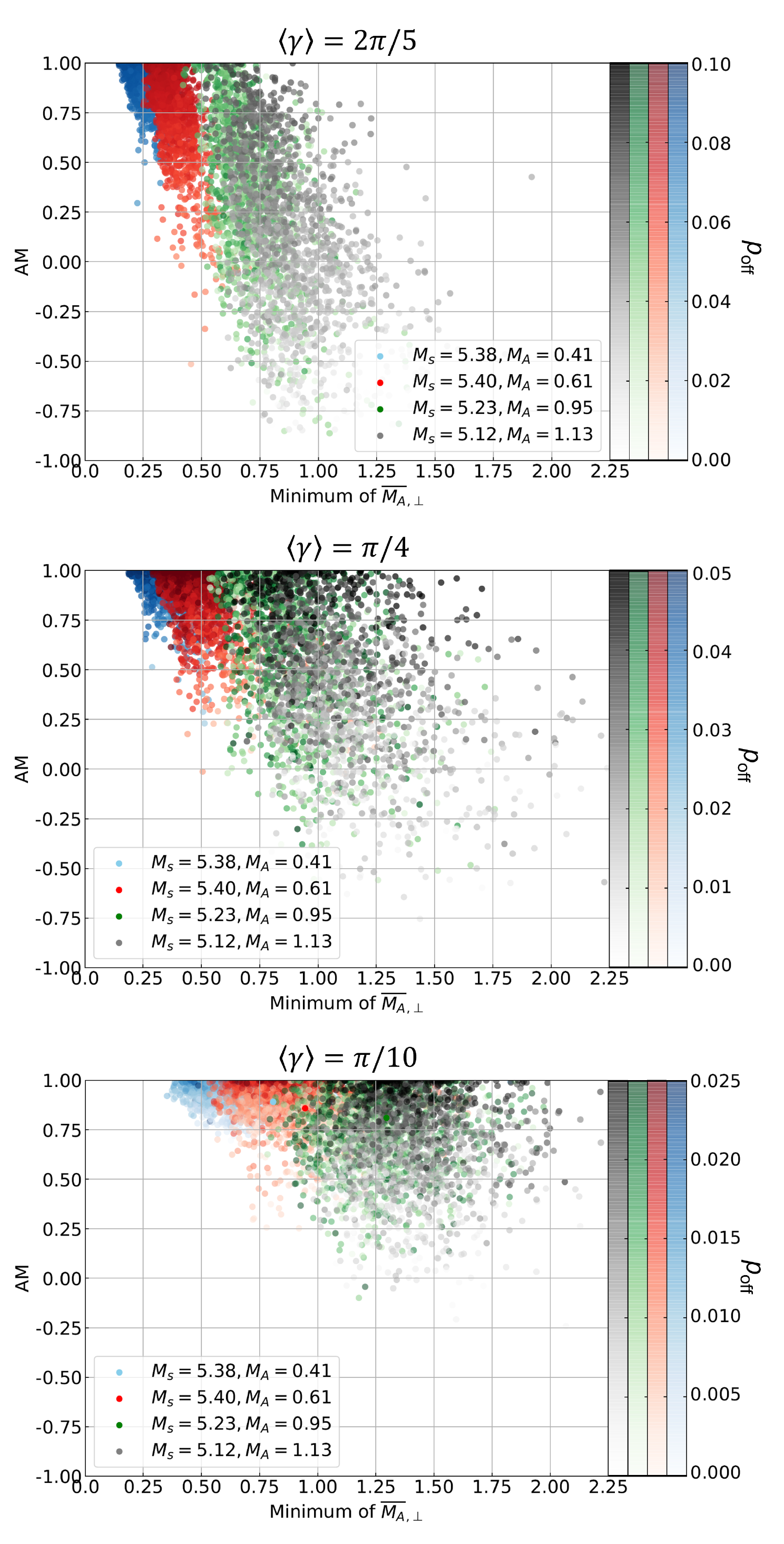}
    \caption{Scatter plots of minimum ${\overline{M_{\rm A}}_{,\bot}}$ and AM (of $\langle\gamma\rangle_{\rm off}$ and $\langle\gamma\rangle_{\rm sub}$). Minimum $\overline{M_{\rm A}}_{,\bot}$ and AM are calculated for each $22\times22$ cell$^2$ sub-region. Color indicates the polarization fraction $p_{\rm off}$ corresponding to minimum ${\overline{M_{\rm A}}_{,\bot}}$. $\langle\gamma\rangle$ denotes the global mean inclination angle of the full simulation box.}
    \label{fig:mamin}
\end{figure}

Fig.~\ref{fig:3D} present an example of the inclination angles measured for sixteen sub-regions, whose size is $198\times198$ cell$^2$. For simplicity, the sub-region is defined as a square, and we refer to its size using the length scale in the following. Each vector is constructed by the POS magnetic field's position angle (i.e., $\psi+\pi/2$) inferred from Stokes parameters (see \S~\ref{sec:theory}) and the inclination angle of either measured $\langle\gamma\rangle_{\rm off}$ or actual $\langle\gamma\rangle_{\rm sub}$ of that sub-region. As we see, globally, the simulation has inclination $\langle\gamma\rangle=\pi/4$ and the POS magnetic field is along the $x$-axis. While the magnetic field's orientation exhibits slight variation for each sub-region, the measured inclination angles agree well with the actual angles. 

Moreover, we test the accuracy of $\langle\gamma\rangle_{\rm off}$ with various sub-region sizes. The global agreement of $\langle\gamma\rangle_{\rm off}$ and $\langle\gamma\rangle_{\rm}$ is quantified by the AM (see \S~\ref{sec:data}). As shown in Fig.~\ref{fig:sb}, in general, the AM increases for a large sub-region size. This can be easily understood as a large sub-region means the probability of finding out $\min\{\overline{M_{\rm A}}_{,\bot}^2\}\ll1$ increases. Therefore, the estimation for a large sub-region is always more accurate. Also, we note that in the  super-Alfv\'enic case (i.e., $M_{\rm A}=1.13)$, the increment of AM at a large sub-region is more significant than the sub-Alfv\'enic case. This indicates that the accuracy of the estimated inclination angle mainly depends on the condition that whether there exists a position with $\overline{M_{\rm A}}_{,\bot}^2\ll1$. As super-Alfv\'enic turbulence has significant magnetic field fluctuations, it is possible that in several positions, the local physical condition becomes sub-Alfv\'enic. Consequently, the probability of finding out a position with $\overline{M_{\rm A}}_{,\bot}^2\ll1$ increases in a large sub-region.


In addition, we notice that the estimation of $\langle\gamma\rangle_{\rm off}$ is more accurate when the actual mean inclination angle $\langle\gamma\rangle$ is small. Intuitively this disagrees with our theoretical consideration that large $\langle\gamma\rangle$ suggests a small value of $\overline{M_{\rm A}}_{,\bot}$, which better constrains $\langle\gamma\rangle_{\rm off}$. However, the crucial term in determining $\langle\gamma\rangle_{\rm off}$ is $\overline{M_{\rm A}}_{,\bot}^2\sin^2\overline{\gamma}$ instead of $\overline{M_{\rm A}}_{,\bot}^2$ (see Eq.~\ref{eq:p}). The choice of using $\overline{M_{\rm A}}_{,\bot}^2\ll1$ is based on the fact it is the only achievable variable in observation. For a given $\overline{M_{\rm A}}$ value, a small inclination angle significantly and non-linearly reduces the value of $\overline{M_{\rm A}}_{,\bot}^2\sin^2\overline{\gamma}$. For instance, $\sin^2(\pi/10)$ is  one order of magnitude smaller than $\sin^2(2\pi/5)$. Therefore, $\sin^2\overline{\gamma}$ becomes the leading factor when the mean inclination angle is small and consequently, Eq.~\ref{eq.17} is better constrained with a small inclination angle. 

Fig.~\ref{fig:mamin} presents the relation of $\min\{\overline{M_{\rm A}}_{,\bot}\}$ and the AM (of $\langle\gamma\rangle_{\rm off}$ and $\langle\gamma\rangle_{\rm sub}$) calculated for each $22\times22$ cell$^2$ sub-region. $\min\{\overline{M_{\rm A}}_{,\bot}\}$ is the minimum value of $\overline{M_{\rm A}}_{,\bot}$ within one sub-region. The sub-region $22\times22$ cell$^2$ cells guarantees sufficient samples for  characterizing overall statistical properties. First of all, as we expected, a small value of $\min\{\overline{M_{\rm A}}_{,\bot}\}$ is associated with large AM, i.e., high accuracy, as well large polarization fraction. 

For the case of $\langle\gamma\rangle=2\pi/5$, the AM starts dropping to negative when $\min\{\overline{M_{\rm A}}_{,\bot}\}> 0.50$. In this situation, the contribution from $\overline{M_{\rm A}}_{,\bot}^2\sin^2\overline{\gamma}$ is not negligible so that the assumption of Eq.~\ref{eq.17} breaks. A smaller inclination angle $\langle\gamma\rangle=\pi/10$ shifts $\min\{\overline{M_{\rm A}}_{,\bot}\}$ to larger value and increases AM. For such a small $\langle\gamma\rangle$, $\overline{M_{\rm A}}_{,\bot}$ is less important in determining $\langle\gamma\rangle_{\rm off}$. In observation, $\langle\gamma\rangle_{\rm off}$ can be obtained from Eq.~\ref{eq.17} targeting on the full cloud. Once the value of $\langle\gamma\rangle_{\rm off}$ is available, the sub-region size can be selected accordingly. One should use a pretty large size when both $\min\{\overline{M_{\rm A}}_{,\bot}\}>0.5$ and $\langle\gamma\rangle_{\rm off}$ is large (for instance, $\langle\gamma\rangle_{\rm off}>\pi/4$). Otherwise, if $\langle\gamma\rangle_{\rm off}$ is small, the restriction on $\min\{\overline{M_{\rm A}}_{,\bot}\}$ and sub-region size can be released.
Note that in real observation that $\overline{M_{\rm A}}_{,\bot}$’s estimation also has uncertainty. Therefore, unlike our numerical results of using the $\min\{\overline{M_{\rm A}}_{,\bot}\}$ in Fig.~\ref{fig:mamin}, it is better to search for a number of reference positions satisfying $\overline{M_{\rm A}}_{,\bot}^2\ll1$ and check the corresponding polarization fraction and inclination angle's variation.

\section{Discussion}
\label{sec:dis}
\subsection{Assumption and uncertainty}
In this work, we propose a method, i.e., the Polarization Fraction Analysis (PFA), to estimate the mean inclination angle of a cloud $\langle\gamma\rangle$. This method is based on several important assumptions. First of all, we assume the existence of a mean magnetic field and the mean field's variation along the LOS is small. This is typically valid for cloud-scale, clump-scale, and core-scale objects. We generally call these objects clouds in the paper. Secondly, we assume the intrinsic polarization fraction $p_0$ is constant throughout a cloud. This implicitly requires that dust grains’ properties, i.e., emissivity, temperature, etc., are homogeneous within the cloud. The other important assumptions related to incompressible MHD turbulence and uncertainty from the underestimation of $p_{\rm max}$ are discussed below.

\subsubsection{Incompressible and compressible MHD turbulence}
Our proposed method accommodates magnetic field fluctuations along the LOS considering incompressible MHD turbulence. This consideration builds up a simple magnetic field model, i.e., the fluctuation is dominantly along the direction perpendicular to the mean magnetic field.

The existence of a mean magnetic field implicitly assumes the MHD turbulence is sub- or trans-Alfv\'enic. Super-Alfv\'enic MHD turbulence is typically isotropic, and a mean field cannot be well defined. Nevertheless, as turbulence cascades to small scales, the importance of magnetic backreaction gets stronger. Eventually, at and below the scale $l_{\rm A} = L_{\rm inj}M_{\rm A}^{-3}$, the turbulent velocity becomes equal to the Alfv\'en velocity, and the turbulence becomes anisotropic \citep{2006ApJ...645L..25L}. Therefore, for the application to a globally super-Alfv\'enic cloud, it is necessary that the telescope can resolve  the scale smaller than $l_{\rm A}$.

Moreover, in a real scenario, ISM turbulence consists of compressible fast and slow modes. Nevertheless, both slow and fast modes in low-$\beta$ plasma are highly anisotropic \citep{2003MNRAS.345..325C,2017MNRAS.464.3617K}, i.e., the most significant fluctuations appear in the perpendicular direction. Here $\beta=2(\frac{M_{\rm A}}{M_s})^2$ is plasma's compressibility. It suggests that in low-$\beta$ molecular clouds, our assumption about perpendicular magnetic field fluctuation is still valid in compressible turbulence. This is also numerically confirmed in Fig.~\ref{fig:hist}. 

The slow mode in high-$\beta$ plasma is similar to the pseudo-Alfv\'en mode in the incompressible regime, while the high-$\beta$ fast mode is a purely compressible mode with an isotropic power spectrum. Although the maximum energy fraction of fast mode is only $\sim20\%$ \citep{2021arXiv211115066H}, an additional consideration is probably necessary to deal with the isotropic fast mode in high-$\beta$ MHD turbulence. 

In addition, incompressible MHD turbulence implicitly means the absence of density fluctuations that are not negligible in observation. However, as shown in Fig.~\ref{fig:pmax}, the leading factor of depolarization is inclination angle, rather than density fluctuation and magnetic field strength's fluctuation. This suggests that the density fluctuation's role is insignificant even in supersonic turbulence. 
\subsubsection{Underestimation of $p_{\rm max}$}
$p_0$ is a key parameter in deriving the inclination angle. \citet{2019MNRAS.485.3499C} argued that $p_0$ depends purely on the intrinsic properties of dust grain and can be calculated from $p_0=3p_{\rm max}/(3+p_{\rm max})$, where $p_{\rm max}$ is the ideal maximum polarization fraction corresponding to local inclination angle $\sim90^\circ$. The observed $p_{\rm max}$ in observation, however, might not satisfy the condition. Consequently, the observed $p_{\rm max}$ is underestimated compared with the ideal value. As shown in Fig.~\ref{fig:pmax}, this underestimation is more significant when $\langle\gamma\rangle<45^\circ$ and it introduces uncertainty to the estimated inclination angle. 

All the assumptions mentioned above can cause systematic uncertainties in the PFA. As we numerically studies in Fig.~\ref{fig:sigma}, the total systematic uncertainty ranges from 0 to $20^\circ$ with a median value $\le10^\circ$. 

Moreover, the estimated inclination angle is in the range of [0, $90^\circ$] (see Eq.~\ref{eq.17}). It does not distinguish whether magnetic field is oriented in the first and third quadrants, as defined in Fig.~\ref{fig:1}, or in the second and fourth quadrants. This degeneracy potentially can be solved by the recent development of Faraday rotation method \citep{2022arXiv220104718T}. 



\subsection{Mapping the POS $M_{\rm A}$ distribution}
The proposed method of probing three-dimensional magnetic fields requires maps of observed polarization fraction and $\overline{M_{\rm A}}_{,\bot}$ distribution. We list several approaches of getting $\overline{M_{\rm A}}_{,\bot}$ here.
 
The first way is using the polarization measurement. For instance, \citet{2008ApJ...679..537F} suggested a generalization of the Davis–Chandrasekhar–Fermi method \citep{1951PhRv...81..890D,1953ApJ...118..113C} to obtain the $\overline{M_{\rm A}}_{,\bot}$ by $\tan\delta\theta\sim\overline{M_{\rm A}}_{,\bot}$. Here $\delta\theta$ is the dispersion of polarization angles.

Also, the $\overline{M_{\rm A}}_{,\bot}$ can be calculated from the polarization fraction using the relation $\sigma_{\rm pol\%}\sim\overline{M_{\rm A}}_{,\bot}^2$ \citep{Lazarian18}, where $\sigma_{\rm pol\%}$ is the dispersion of polarization fraction. Although the measurement of $\delta\theta$ or $\sigma_{\rm pol\%}$ over a region reduces the observation's resolution, once the $\overline{M_{\rm A}}_{,\bot}$ distribution is available, as presented in \cite{Lazarian18}, \citet{2021ApJ...913...85H} and \citet{2021MNRAS.tmp.3119L}, one can access the three-dimensional magnetic field using our proposed PFA.

The velocity gradient technique (VGT; \citealt{GL17,LY18a,HYL18}) and the structure-function analysis (SFA; \citealt{HXL21,XH21,HLX21a}) are other two approaches of getting $\overline{M_{\rm A}}_{,\bot}$. The VGT relies on the fact that velocity gradient's dispersion is small in a strongly magnetized medium, but becomes large in weak magnetized medium. The relation of velocity gradient's dispersion and $\overline{M_{\rm A}}_{,\bot}$ is given in \cite{Lazarian18}.


The SFA estimates $\overline{M_{\rm A}}_{,\bot}$ from the ratio of velocity fluctuations perpendicular and parallel to the POS magnetic field. Its foundation is also MHD turbulence's anisotropy, which suggests that the maximum velocity fluctuation appears in the direction perpendicular to the magnetic field, but the minimum appears in the parallel direction. Their ratio is positively proportional to $(\overline{M_{\rm A}}_{,\bot})^{-4/3}$. 
 
Moreover, the VGT and SFA potentially contain the necessary information for getting pixelized distributions of total magnetic field strength and inclination angle from the Eq.~\ref{eq:ma}. The dilemma of Eq.~\ref{eq:ma} is that we need sufficient samples to constrain turbulence's property, which does not appear in a single data point of dust polarization. However, the Doppler-shifted lines used by the VGT or SFA usually has a higher resolution than polarization measurement. For example, the CO (1-0) emission line observed with the Green Bank Telescope achieves a beam resolution $\sim8''$. If one selects a sub-region size smaller than $80\times80$ cell$^2$, the measured turbulence's property by the VGT or SFA for each sub-region would have resolution $\sim10'$, which is comparable with the Planck polarization measurement. This information, therefore, could be implemented in Planck polarization to obtain local magnetic field strength and inclination angle.

\subsection{Comparison with Other Methods}
\citet{2019MNRAS.485.3499C} proposed a method to calculate the inclination angle of the magnetic field. Assuming idealistic and homogeneous physical conditions, i.e., magnetic field's fluctuations are negligible, $p_0$ is constant across the cloud, and dust grains' properties are the same, their method calculates the inclination angle distribution using the local polarization fraction (see Eq.~\ref{eq:chen}). However, the assumption holds only for the strongly sub-Alfv\'enic case, while molecular clouds are typically trans-Alfv\'enic or super-Alfv\'enic \citep{2016ApJ...832..143F,Hu19a,2021ApJ...913...85H,2021MNRAS.tmp.3119L,2022arXiv220512084T}. The systematic uncertainty of their method in trans-Alfv\'enic or super-Alfv\'enic regimes ranges from  $\sim3^\circ$ to $\sim35^\circ$. 


In this work, we take into account the fluctuation of the magnetic field due to anisotropic MHD turbulence. We show that the local polarization fraction, in this case, depends on not only the inclination angle but also the magnetic field fluctuation. The fluctuation amplifies the depolarization effect. Consequently, the local polarization fraction does not accurately characterize the inclination angle using Eq.~\ref{eq:chen}. We propose and demonstrate that the polarization fraction in the reference position with $\overline{M_{\rm A}}_{,\bot}^2\ll1$ is determined by the mean inclination angle over a region of interest since the contribution from the fluctuation is insignificant there. The mean inclination angle thus can be calculated (see Eq.~\ref{eq.17}) once the distribution of $\overline{M_{\rm A}}_{,\bot}^2$ is available. In particular, our method is applicable to molecular clouds. Because trans-Alfv\'enic or super-Alfv\'enic clouds raise significant $\overline{M_{\rm A}}_{,\bot}$ fluctuations, one can easily find a position corresponding to $\overline{M_{\rm A}}_{,\bot}^2\ll1$ by searching for sufficient samples.

Another two methods of tracing three-dimensional magnetic fields were proposed by \citet{HXL21} and \citet{HLX21a}. The two methods are based on MHD turbulence's anisotropic property, i.e., the maximum velocity fluctuation appears in the direction perpendicular to the magnetic field. Consequently, by measuring the three-dimensional velocity fluctuations of young stellar objects, which are accessible via the Gaia survey \citep{2016A&A...595A...1G,2018A&A...616A...1G,2021ApJ...907L..40H,2022arXiv220500012H}, one can find the three-dimensional magnetic fields \citep{HXL21}. \citet{HLX21a}, on the other hand, proposed to measure the velocity fluctuations using Doppler-shifted emission lines. It was shown that the ratio of maximum and minimum fluctuations within a given velocity channel is correlated with the velocity channel width, total Alfv\'en Mach number, and the inclination angle. Therefore, by varying the channel widths used for calculating the ratio, one can solve the $M_{\rm A}$ and inclination angle simultaneously.

\section{Summary}
\label{sec:con}
Dust polarization is one of the most important ways to trace the magnetic fields in ISM. We propose a new method, i.e., the PFA, to trace three-dimensional magnetic fields using the observed polarization fraction of polarized dust emission and the distribution of the POS Alfv\'en Mach number. This method mainly assumes the existence of a mean magnetic field in a physically homogeneous cloud and magnetic field fluctuations arise from anisotropic MHD turbulence. 
We summarize as follows:
\begin{enumerate}
    \item We numerically confirm that magnetic fluctuation of compressible turbulence dominantly appears in the direction perpendicular to the mean magnetic field.
    \item We find inclination angle is the primary agent for depolarization. Fluctuations of magnetic field strength and density have an insignificant contribution.
    \item We analytically propose and numerically confirm that the polarization fraction corresponding to $\overline{M_{\rm A}}_{,\bot}^2\ll 1$ can characterize the mean inclination angle. 
    \item We test the PFA using 3D compressible MHD turbulence simulations and show that it is applicable to sub-Alfv\'enic, trans-Alfv\'enic, and moderately supers-Alfv\'enic clouds with $0.4\lesssim M_{\rm A}\lesssim1.2$. 
    \item We numerically find the PFA has systematic uncertainty ranging from 0 to $20^\circ$ with a median value $\le10^\circ$.
\end{enumerate}

\section*{Acknowledgements}
Y.H. and A.L.acknowledges the support of NASA ATP AAH7546. Financial support for this work was provided by NASA through award 09\_0231 issued by the Universities Space Research Association, Inc. (USRA). We thank the reviewer for numerous suggestions for improving the manuscript. We acknowledge the allocation of computer time by the Center for High Throughput Computing (CHTC) at the University of Wisconsin-Madison.

\section*{Data Availability}
The data underlying this article will be shared on reasonable request to the corresponding author. 


\bibliographystyle{mnras}
\bibliography{example} 




\appendix
\section{3D Turbulent velocity estimated from emission line}
\label{appendix}
\begin{figure}
\centering
	\includegraphics[width=1.0\linewidth]{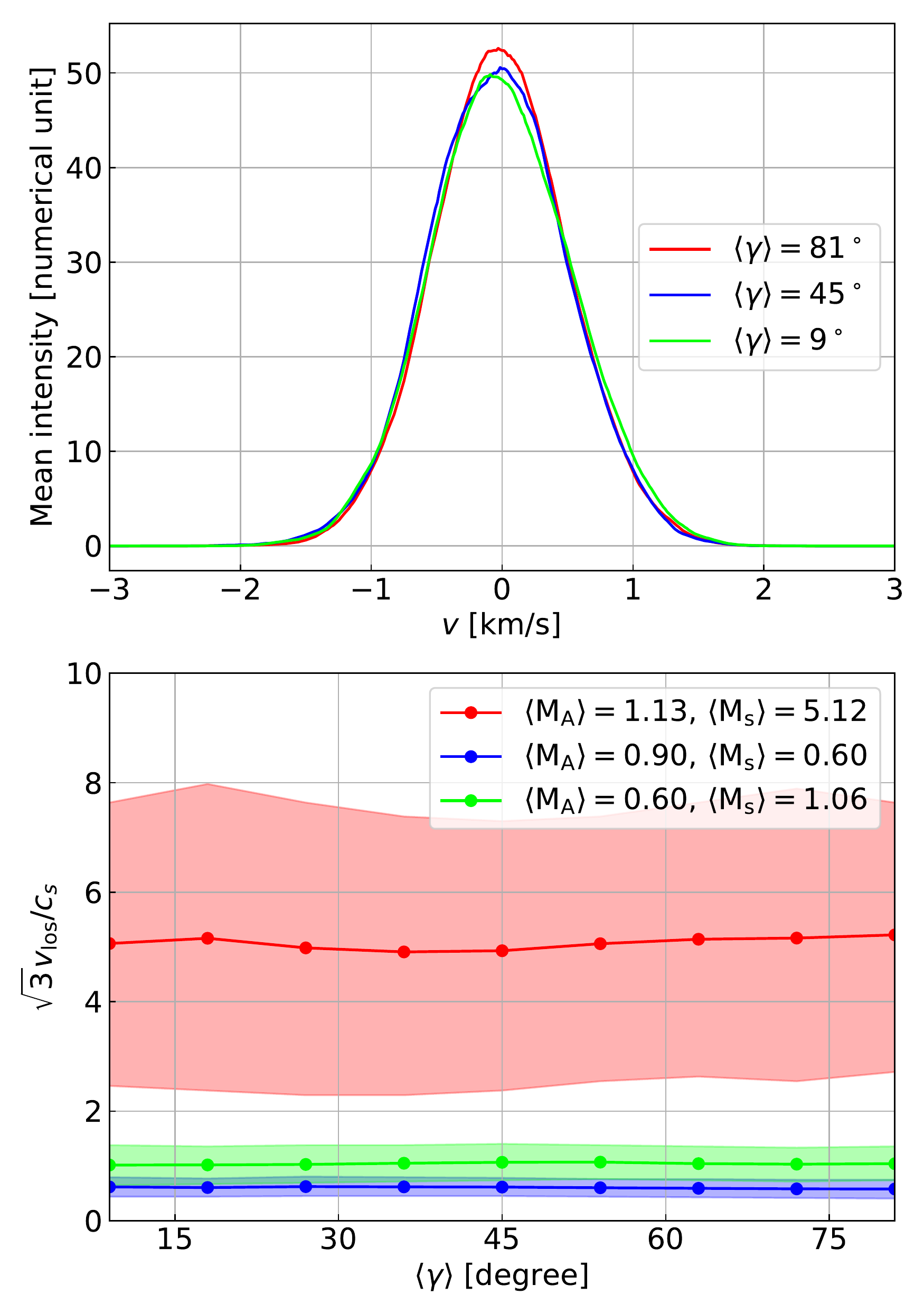}
    \caption{\textbf{Top}: synthetic spectra generated from the simulation $\langle M_{\rm A}\rangle=1.13, \langle M_{\rm s}\rangle=5.12$ with various $\langle\gamma\rangle$. \textbf{Bottom:} the correlation of  median $\sqrt{3}v_{\rm los}/c_s$ and mean inclination angle $\langle\gamma\rangle$. $c_s\approx0.192$ is the speed of sound in numerical unit. The shadowed area indicates the ranges of one-sigma level.}
    \label{fig:vlos}
\end{figure}

To find the distribution of $\overline {M_{\rm A}}_{,\bot}$, knowledge of turbulent velocity $v_l=(l/L_{\rm inj})^{1/3}v_{\rm inj}$ at scale $l$ is required, assuming Kolmogorov-type turbulence. Here $L_{\rm inj}$ is the injection scale and $v_{\rm inj}=\sqrt{3}v_{\rm los}$ is the 3D turbulent velocity at injection scale. This calculation needs the emission line's width $v_{\rm los}$ and isotropic turbulence at the injection scale.

We use three numerical simulations $\langle M_{\rm A}\rangle=1.13, \langle M_{\rm s}\rangle=5.12$, $\langle M_{\rm A}\rangle=0.90, \langle M_{\rm s}\rangle=0.60$, and $\langle M_{\rm A}\rangle=0.60, \langle M_{\rm s}\rangle=1.06$ to test the validity of $v_{\rm inj}=\sqrt{3}v_{\rm los}$ with different $\langle\gamma\rangle$. The simulation setup is the same as the one presented in \S~\ref{sec:data}. 

We extract the LOS velocity and density information from MHD simulations to generate synthetic Position-Position-Velocity (PPV) cubes. Fig.~\ref{fig:vlos} presents the spectra calculated from the PPV cubes generated from the simulation $\langle M_{\rm A}\rangle=1.13, \langle M_{\rm s}\rangle=5.12$ in the conditions of $\langle\gamma\rangle=81^\circ$, $\langle\gamma\rangle=45^\circ$, and $\langle\gamma\rangle=9^\circ$. The spectra are averaged over the full PPV cubes along the $x$ and $y$-directions, i.e., the POS. We ensure that the same number of pixels enter the calculations and that the spectra are calculated in the same interval and with the same bandwidth. We can see the spectral width has insignificant changes. Moreover, we calculate $v_{\rm los}$ from $v_{\rm los}={\rm FWHM}/\sqrt{8\ln{2}}$, where ${\rm FWHM}$ stands for the full width at half maximum. The FWHM is independently calculated for the spectrum in each pixel. The median value of $\sqrt{3}v_{\rm los}/c_s$ gives an estimation of the simulation's intrinsic $\langle M_{s}\rangle$ at injection scale. In Fig.~\ref{fig:vlos}, we can see, in all sub-sonic, trans-sonic, and supersonic conditions, $\sqrt{3}v_{\rm los}/c_s$ has only insignificant variation when $\langle\gamma\rangle$ changes. However, although we expect isotropic turbulence driving in most cases, readers should be careful about the situation of anisotropic driving. Anisotropic driving would result in either an underestimation or overestimation of $\overline {M_{\rm A}}_{,\bot}$ using the relation $v_{\rm inj}=\sqrt{3}v_{\rm los}$.


\bsp	
\label{lastpage}
\end{document}